\documentclass[aps,pra,reprint,floatfix]{revtex4-2}

\usepackage[utf8]{inputenc}
\usepackage[T1]{fontenc}
\usepackage{lmodern}

\usepackage[
    main= american,
    german,
    ]{babel}

\usepackage{amsfonts,amsmath,amssymb,stmaryrd}
\usepackage{physics}
\usepackage{mathrsfs}  
\usepackage{mathtools}
\usepackage{graphicx}
\usepackage{xcolor}

\usepackage{braket}
\usepackage{hyperref}
\usepackage[capitalise]{cleveref}

\usepackage{verbatim}

\usepackage{lipsum}

\usepackage{ulem}

\newcommand{\gib}{g_\text{IB}}

\newcommand{\dx}{\text{d}x}
\newcommand{\dy}{\text{d}y}

\newcommand{\xit}{\tilde{\xi}}

\newcommand{\hc}{\textrm{h.c.}}
\newcommand{\figref}[1]{Fig.~\ref{#1}}
\newcommand{\secref}[1]{Sec.~\ref{#1}}
\renewcommand{\eqref}[1]{eq.~(\ref{#1})}

\newcommand{\dn}{\textrm{dn}}
\newcommand{\cn}{\textrm{cn}}
\newcommand{\cd}{\textrm{cd}}
\newcommand{\sn}{\textrm{sn}}
\newcommand{\am}{\textrm{am}}

\renewcommand{\emph}[1]{\textit{#1}}

\begin{document}
%%%%%%%%%%%%%%%%%%%%%%%%%%%%%%%%%%%%%%%%%%%%%%%%%
\title{Impurities in a trapped 1D Bose gas of arbitrary interaction strength: localization-delocalization transition and absence of self-localization}
%%%%%%%%%%%%%%%%%%%%%%%%%%%%%%%%%%%%%%%%%%%%%%%%%

\author{Dennis Breu}
\author{Eric Vidal Marcos}
\author{Martin Will}
\author{Michael Fleischhauer}
\affiliation{Department of Physics and Research Center OPTIMAS, University of Kaiserslautern-Landau, 67663 Kaiserslautern, Germany}

%%%%%%%%%%%%%%%%%%%%%%%%%%%%%%%%%%%%%%%%%%%%%%%%%

\begin{abstract}
We discuss impurities in a  one-dimensional Bose gas with arbitrary boson-boson and boson-impurity interactions. To fully account for quantum effects,  we employ numerical simulations based on the density-matrix renormalization group (DMRG) and - in the regime of strong boson-boson interactions - the mapping to weakly interacting fermions. 
A mean-field description of the Bose polaron based on coupled Gross-Pitaevski -- Schr\"odinger equations predicts the existence
of a self-localized polaron. We here show that such a solution does not exist  and is an artifact of the underlying decoupling approximation. To this end we consider a
mobile impurity in a box potential.  Our work demonstrates that
correlations between the impurity position and the bosons are important even in the limit where mean-field approaches are expected to work well. 
Furthermore we derive analytical approximations for the energy of a single polaron formed by a heavy impurity for arbitrary interaction strengths and large
but finite boson-boson couplings which accurately reproduce DMRG results. This demonstrates that the polaron problem of a heavy impurity
in a 1D Bose gas  can be accurately approximated by a proper mean-field description plus a linearized treatment of quantum fluctuations for arbitrary boson-boson and impurity-boson couplings.
Finally we  determine the polaron-polaron interaction potential $V(r)$ in Born-Oppenheimer approximation for small and intermediate distances $r$, which in the Tonks gas limit 
is oscillatory due to Friedel oscillations in the Bose gas. 
\end{abstract}

\date{\today}
\maketitle
 
%%%%%%%%%%%%%%%%%%%%%%%%%%%
%%%%%%%%%%%%%%%%%%%%%%%%%%%
\section{Introduction}
%%%%%%%%%%%%%%%%%%%%%%%%%%%
%%%%%%%%%%%%%%%%%%%%%%%%%%%

Quasi particles formed by quantum impurities immersed in a many-body environment play a key role for the understanding of transport phenomena \cite{Alexandrov2010}.
Furthermore the interaction of quasi particles mediated by the host medium 
forms the basis of many important many-body phenomena in condensed-matter physics
\cite{Ruderman1954,Kasuya1956,Yosida1957,Cooper1956}.
In recent years the possibility to experimentally investigate impurities in 
quantum fluids, such as Bose-Einstein condensates (BEC) of atoms has renewed the interest in these quasi-particles, called Bose polarons. 
An important difference between Bose polarons in ultra-cold quantum gases and
the Landau-Pekar polaron \cite{Landau1933,Pekar1946} introduced to model electrons interacting with the lattice vibrations of a solid is the large compressibility of the BEC. As a consequence  a common theoretical model similar to 
 the Fr\"ohlich model in solids \cite{Frohlich1954}, which describes the polaron as interaction of the impurity with phonon excitations, is only suitable for weak boson-boson and weak impurity-boson interactions \cite{Grusdt2017b}. In this limit the condensate can be considered undepleted and the role of lattice vibrations is taken over by the Bogoliubov phonons. 

For stronger interactions with the impurity the back-action to the condensate needs to be taken into account, while keeping correlations between impurity position and bosons. For translational invariant systems this can be done by means of a Lee-Low-Pines (LLP) transformation \cite{Lee1953}, which decouples the impurity motion from the many-body problem of interacting bosons. For weak boson-boson interactions $g$ the latter can then be treated rather accurately in a mean-field description for arbitrary strength of the
impurity-boson couplings $g_\textrm{IB}$ \cite{Volosniev2017,Jager2020,Will2021,will2023dynamics,Panochko2019}.

In an inhomogeneous Bose gas and for repulsive impurity-boson interactions, the polaron can localize in density minima of the Bose gas
and, similarly to phase separation in multi-component condensates \cite{timmermans1998phase}, a localization-delocalization transition 
can occur. Due to the absence of translational invariance the LLP decoupling does not work here, however. Instead another mean-field ansatz is often used, which in addition neglects however correlations between impurity position and bosons and results in a coupled Gross-Pitaevski and Schr\"odinger equation for the condensate and the impurity 
respectively \cite{lee1992polarons,cucchietti2006strong,sacha2006self,kalas2006interaction,bruderer2008self,blinova2013two}. 
This decoupling mean-field approach (DMF) also predicts a localization transition to occur in translationally invariant systems.
In analogy to impurities in liquid $^4$He \cite{brewer1966quantum,hernandez1991electron}, a single atom can become \emph{self-trapped} in a distortion of the condensate created by the impurity 
\cite{lee1992polarons,cucchietti2006strong,sacha2006self,kalas2006interaction,bruderer2008self,blinova2013two} thereby spontaneously breaking translational invariance. In 2D and 3D a critical interaction strength is needed for  such a self-trapped polaron to exist, but in 1D  within the DMF approach an arbitrarily small $\gib$ is sufficient \cite{bruderer2008self,blinova2013two}. 
Recently it was shown in Ref.\cite{zschetzsche2024suppression} that correlations between the center-of-mass motion of the impurity and the bosons suppress this self-localization transition and
solutions with finite localization length only survive for large $\gib$.

%
%%%%%%%%%%%%%%%%%%%%%%%%%%%%%%%%%%%%%%%%%%%%%%
\begin{figure}[htb]
\centering
\includegraphics[width=0.8\columnwidth]{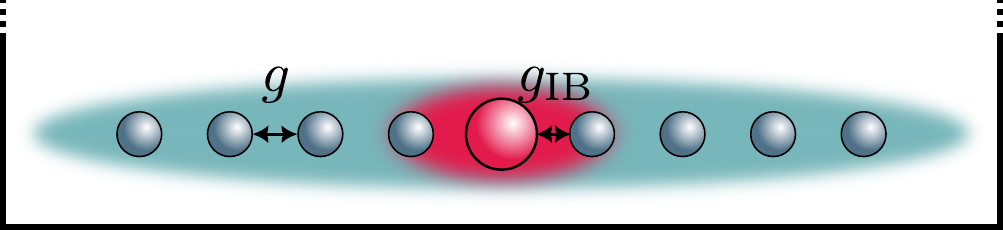}
\caption{Mobile quantum impurity in a one-dimensional, interacting Bose gas, trapped in a box potential. Boson-boson interaction is characterized by coupling strength $g$ and the impurity interacts with the Bose gas with strength $g_\textrm{IB}$.
} \label{fig:system} 
\end{figure}
%%%%%%%%%%%%%%%%%%%%%%%%%%%%%%%%%%%%%%%%%%%%%
%

In the present paper we analyze the role of quantum fluctuations to  
the self-trapping. 
To this end we consider a mobile impurity of finite mass $M$ in a one-dimensional condensate and employ 
numerical simulations based on the density-matrix renormalization group (DMRG) \cite{schollwock,itensor}. Furthermore we consider a box potential, see \figref{fig:system}, to explicitly break translational invariance. For such a system a DMF approach predicts a reentry transition from a
\emph{self-localized} polaron at the trap center to a delocalized polaron and eventually to one localized at the potential edge upon changing $\gib$.
We numerically determine the ground-state phase diagram taking quantum fluctuations into account and observe instead only a phase transition between a fully \emph{delocalized} phase for small values of the impurity-boson coupling $\gib$
% where the impurity is distributed over the whole trap 
%, i.e. with a wavefunction centered at the middle of the trap and a spatial width proportional to the trap size, 
and a localized phase near one of the edges. Our simulations show that there is no self-trapped solution in the full quantum problem even for very large impurity-boson couplings for which a finite localization length was obtained in Ref.\cite{zschetzsche2024suppression}. 

In \cite{Jager2020,Will2021,will2023dynamics} we have shown that for a heavy impurity with mass ratios $M/m > {\cal O}(1)$ in a translational invariant, \emph{weakly interacting} 1D Bose gas, a  mean-field approach in a LLP frame gives very accurate predictions for most properties of the polaron, which can further be improved by adding quantum fluctuations perturbatively. We here 
show that this is also true for a stronly-interacting 1D Bose gas. To this end
we  calculate the polaron energy as function of $\gib$ for \emph{arbitray} boson-boson interaction strength ranging from the Bogoliubov regime to the Tonks-gas regime of (nearly) impenetrable bosons. Making use of the mapping between strongly interacting bosons and weakly interacting fermions, we derive an analytical approximation of the polaron energy in the latter regime which agrees very well with the numerical DMRG data.

Finally we calculate the effective
interaction potential between two heavy impurities in Born-Oppenheimer approximation for short distances. In a weakly interacting Bose gas the potential is monotonous and has a linear slope for small distances \cite{Will2021}. In the  limit of (nearly) impenetrable bosons the potential is modified
by Friedel-like oscillations, whose long-range behavior has been derived in \cite{recati2005casimir,fuchs2007oscillating} using a low-energy approximation. We here extend this approach to also capture the short-distance behavior, relevant for bi-polaron bound states.

The paper is organized as follows: In \secref{sec:model} we introduce the model. 
We discuss the problem of self-localization of a finite-mass impurity in a one-dimensional Bose gas in a box potential in \secref{sec:localization}. 
Then we discuss a mean-field approach to 
single polaron formed by a heavy impurity in a strongly interacting Bose gas in \secref{sec:polaron}. Finally we derive the short-range
interaction potential between two heavy polarons in Born-Oppenheimer approximation
in \secref{sec:bipolaron}.

%%%%%%%%%%%%%%%%%%%%%%%%%%%
%%%%%%%%%%%%%%%%%%%%%%%%%%%
\section{Model}
\label{sec:model}
%%%%%%%%%%%%%%%%%%%%%%%%%%%
%%%%%%%%%%%%%%%%%%%%%%%%%%%

%%%%%%%%%%%%%%%%%%%%%%%%%%%
%\subsection{Model}
%%%%%%%%%%%%%%%%%%%%%%%%%%%

The Hamiltonian describing a single quantum impurity in an interacting 1D Bose gas in a box potential of length $L$ 
has the form ($\hbar =1$)
\begin{align}
    &\hat{H}=\frac{\hat{p}^2}{2M}+\label{eq:Hamiltonian}\\
    & +\int_{-L/2}^{L/2}\!\!\!\! \dx\, \hat{\phi}^\dagger(x)\left[-\frac{\partial^2_x}{2m}+\frac{g}{2}\hat{\phi}^\dagger(x)\hat{\phi}(x)+\gib\delta(x-\hat{r})\right]\hat{\phi}(x),
    \nonumber
\end{align}
where $m$ ($M$) is the boson (impurity) mass and $g$ ($\gib$) the strength of the bose-bose (bose-impurity) $s$-wave interaction. 
$\hat{p}$ and $\hat{r}$ represent the momentum and position operators of the impurity in first quantization. $\hat{\phi}(x)$ is the field operator for the bosons. The bosons and the impurity are trapped in an infinitely deep box potential, described by open boundary conditions at $x=\pm L/2$.
To quantify the internal interaction strength of the 1D Bose gas we use the unitless Tonks parameter
\begin{align}
    \gamma=\frac{gm}{n_0},
\end{align}
where $n_0$ is the mean particle density of bosons in the system. The Tonks parameter gives a relation between the interaction ($\propto gn_0^2$) and the kinetic energy ($\propto n_0^{3}/m$) of the bosons. A large value of $\gamma$ corresponds to strong boson-boson interactions. For $\gamma\ll 1$ and below a critical temperature a quasi condensate forms. (True condensation is not possible due to the Mermin-Wagner-Hohenberg theorem \cite{1D_2D_bec1,1D_2D_bec2}.) For $\gamma\rightarrow\infty$ the bosons become impenetrable hard-core bosons and can be mapped to free fermions \cite{Girardeau1960}.

% \begin{align}
%        \hat{H} &= \int\! dx\, \hat\psi^\dagger(x) \left(-\frac{\partial_x^2}{2 M} +g_\textrm{IB} \hat\phi^\dagger \hat\phi(x) +V_\textrm{I}(x)\right) \hat\psi(x)\\
%        & \quad + \int\! \dx\, \hat{\phi}^\dagger(x)\left(-\frac{\partial^2_x}{2m}+\frac{g}{2}\hat{\phi}^\dagger(x)\hat{\phi}(x)+V(x)\right)\hat{\phi}(x)\nonumber
 %   \end{align}

 %%%%%%%%%%%%%%%%%%%%%%%%%%
 %\subsection{DMRG simulation of quantum impurity in a 1D Bose gas}
 %%%%%%%%%%%%%%%%%%%%%%%%%%

A powerful method to describe the ground-state of quantum many-body systems in 1D is the density matrix renormalization group (DMRG) \cite{schollwock}.
The DMRG method was originally developed for lattice systems. It can also be applied to continuous 
models, however, e.g. using a proper discretization, which we apply here. 
The details of this together with benchmarks are given in the Appendix A.

%%%%%%%%%%%%%%%%%%%%%%%%%%%
%\subsection{Decoupling mean-field (DMF) ansatz}
%%%%%%%%%%%%%%%%%%%%%%%%%%%

The Hamiltonian \eqref{eq:Hamiltonian} contains both the position and momentum operators, $\hat r$ and $\hat p$, of the impurity, which do not commute. This
leads to an entanglement between the motional degrees of freedom of the impurity with the boson field. 
For translationally invariant systems one can solve this problem by a Lee-Low-Pines (LLP) transformation into a co-moving frame \cite{Lee1953}. For weak boson-boson interactions and not too small values of the impurity mass the effective Hamiltonian can then be solved rather accurately in a mean-field approximation  \cite{Jager2020,Will2021}, which due to the LLP transformation  takes impurity-boson correlations into account.

Without translational invariance, e.g. if the Bose gas $(B)$ and / or the impurity $(I)$
are subject to some trapping potentials $V_{B,I}(x)$  a LLP transformation does not work. Here often another mean-field ansatz is used, which neglects correlations between impurity and bosons and results in a coupled Gross-Pitaevski and Schr\"odinger equation for a condensate wavefunction $\phi(x)$ and the impurity wavefunction $\phi_\mathrm{I}(x)$, respectively: 
\begin{align}
    i\partial_t\phi(x,t)=&\biggl(-\frac{\partial_x^2}{2m}+\gib|\phi_\mathrm{I}(x,t)|^2+\nonumber\\ 
    &\quad + g|\phi(x,t)|^2+V_B(x)\biggr)\phi(x,t),\label{eq:GPE-SE}\\
    i\partial_t\phi_\mathrm{I}(x,t)=&\left(-\frac{\partial_x^2}{2M}+\gib|\phi(x,t)|^2+V_I(x)\right)\phi_\mathrm{I}(x,t).\nonumber
\end{align}
We will show here that in contrast to a mean-field theory in the LLP frame, this decoupling approach leads to artifacts.

%%%%%%%%%%%%%%%%%%%%%%%%%%%
%%%%%%%%%%%%%%%%%%%%%%%%%%%
\section{Localization-Delocalization transition in a box potential}
\label{sec:localization}
%%%%%%%%%%%%%%%%%%%%%%%%%%%
%%%%%%%%%%%%%%%%%%%%%%%%%%%

In \cite{Cucchietti2006,Bruderer-EPL2008} it has been argued based on a DMF ansatz, \eqref{eq:GPE-SE},  that in a translational invariant, homogeneous system the impurity wavefunction should self-trap in a co-localized distortion of the BEC for small $\gib$.
The reality of such self-trapped polarons has been subject of discussions. Recently it was shown in  Ref.~\cite{zschetzsche2024suppression} that taking into account leading-order correlations, self trapping 
only occurs above a certain minimum value of $\gib$.

We now show that in a full quantum model self trapping is absent altogether. To this end we numerically investigate a box potential for bosons and impurity $V_B(x)=V_I(x)$. Here we expect a
transition between a delocalized phase, where the impurity is spread out over the whole system, and a localized one, % corresponding to case of phase separation,
where the impurity is localized at one of the two edges of the system.
This is because the energy of an impurity  at the edge increases as $E_\textrm{loc}\sim \sqrt{\gib}$, while the polaron energy in the bulk scales as $E_p\sim \gib$. Thus for $\gib\to 0$ no edge state exists 
which only emerges above a critical value of $\gib$. In addition to this a decoupling mean-field theory predicts another phase transition to a self-localized polaron.
Due to the broken translational invariance in the box potential considered here, such a self-trapping would manifest itself in a probability density of the impurity centered in the middle of the trap with a width independent on the trap size $L$.

In the following we will investigate both localization phenomena for a weakly interacting as well as a strongly interacting Bose gas using DMRG simulations of the full quantum equations and compare them to numerical solutions and analytical approximations of the DMF equations \eqref{eq:GPE-SE}.

%%%%%%%%%%%%%%%%%%%%%%%%%%%
\subsection{Localization-delocalization transition in a box potential}
%%%%%%%%%%%%%%%%%%%%%%%%%%%

\begin{figure}[ht]
\centering
\includegraphics[width=\columnwidth]{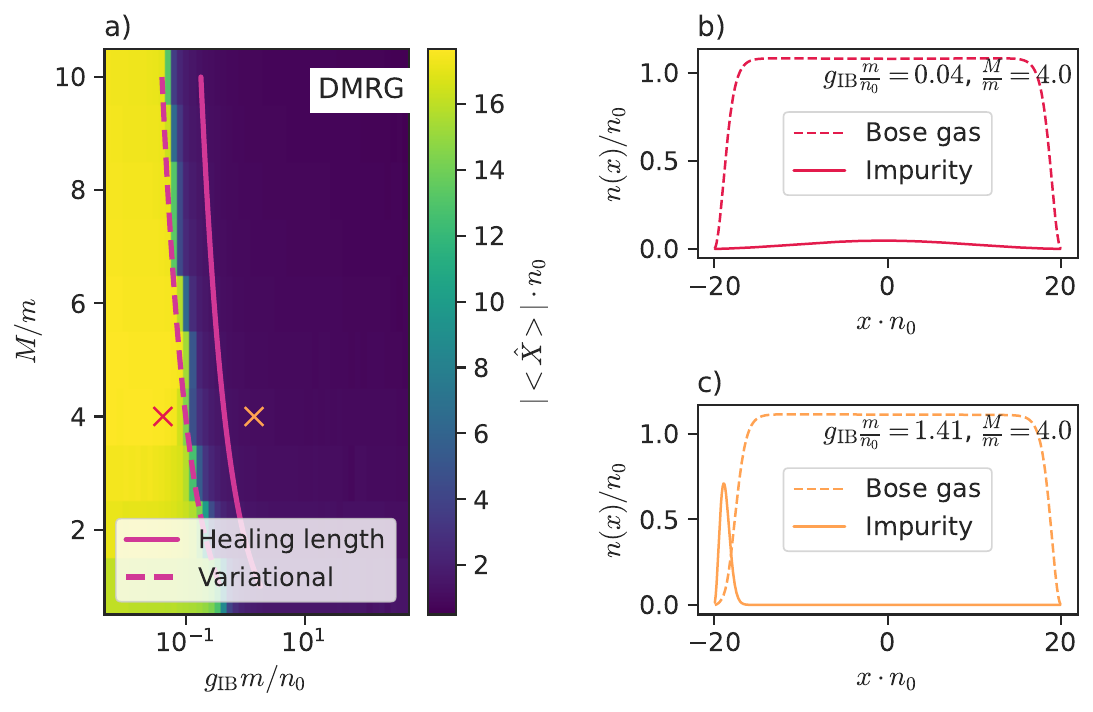}
\caption{a) Localization-delocalization transition of impurity in a box potential in a weakly interacting Bose gas with Tonks parameter $\gamma=0.4$ and $N=40$ particles. Color code describes average position $\langle \hat X\rangle$ of impurity. Solid line is the prediction from \eqref{eq:critical-weak} with $l_0=\xi$. The dashed line gives the prediction with $l_0$ from a variational ansatz. b) and c) density distribution corresponding to the two points in Figure a).} \label{fig:localization-weak} 
\end{figure}

\begin{figure}[ht]
\centering
\includegraphics[width=\columnwidth]{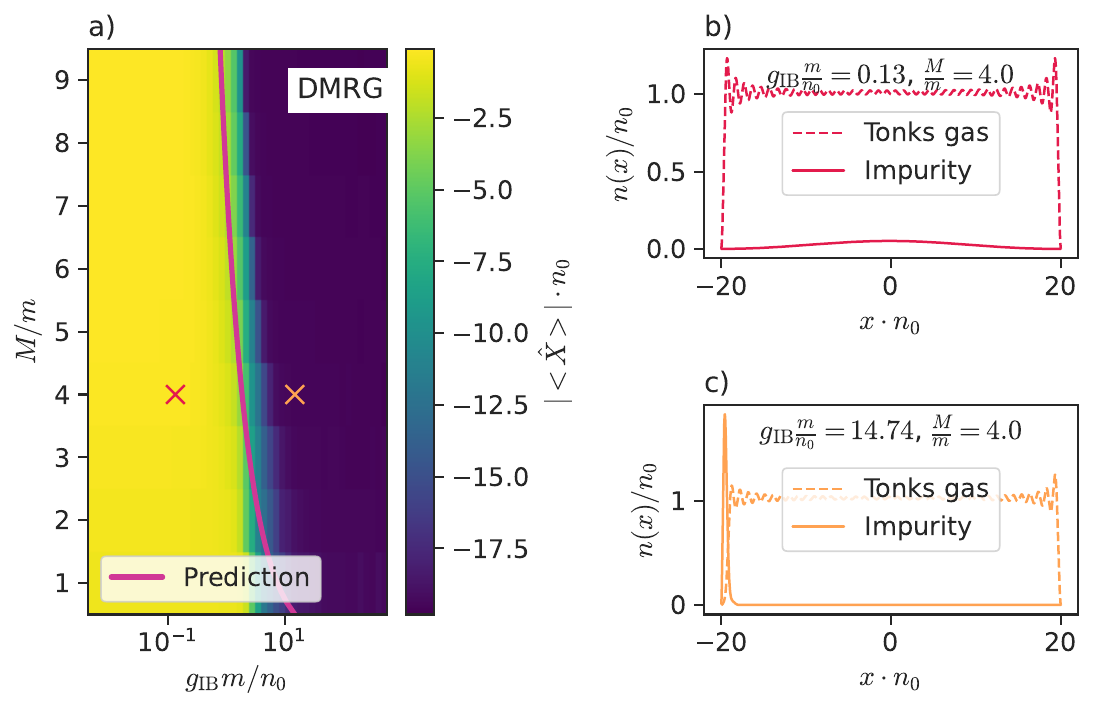}
\caption{a) Localization-delocalization transition of an impurity in a box potential in a Tonks gas ($\gamma=\infty$, $N=40$). Solid line is prediction from \eqref{eq:critical-strong}.
 b) and c) density distribution corresponding to the two points in Figure a).
} \label{fig:localization-strong} 
\end{figure}

We performed DMRG simulations of the ground state of  a mobile impurity in a 1D Bose gas in the box potential for different mass ratios $M/m$ and impurity-boson interaction strengths $\gib$.
Results for $\gamma=0.4$, and  $\gamma=\infty$ are shown in \figref{fig:localization-weak} and \figref{fig:localization-strong}, respectively.
One recognizes in both cases a sharp transition between a phase where the impurity is delocalized over the trap to a phase where it is localized at one of the two edges. (Note that the true ground state is a superposition of localized states at both edges. Since the energy difference between the symmetric and antisymmetric superpositions vanishes exponentially with increasing system size, the DMRG algorithm converges to a solution on one edge only.) The transition point depends on the boson-boson interaction strength.
In the delocalized phase the probability density of the impurity is spread out over the whole system and apart from some corrections at the edges it corresponds to the ground state of the box potential $\sim \cos^2(x\pi/L)$, with $L$ being the length of the box. In the edge-localized phase, on the other hand, the width of the probability distribution becomes independent on system size for large $L$.

%%%%%%%%%%%%%%%%%%%%%%%%%%%%%%%%%%%%%%%%%%%%%%
%%%%%%%%%%%%%%%%%%%%%%%%%%%%%%%%%%%%%%%%%%%%%

A simple estimate for the critical point of the localization-delocalization transition can be obtained as follows:
For weak interactions with a Bose gas of density $n_0$ the energy of a repulsive polaron is given in lowest-order perturbation by
\begin{equation}
    E_p \approx\gib n_0 
\end{equation}
and thus scales linearly in $\gib$. 

On the other hand the energy of an impurity localized at the edge of the condensate can be estimated from the repulsive potential created by the bosons, whose density increases quadratically close to the potential edge:

For a weakly interacting Bose gas (small $\gamma$) the 
characteristic length scale $l_0= \eta \xi $ of the harmonic confinement is the healing length $\xi$ of the Bose condensate. The factor $\eta \simeq {\cal O}(1)$ accounts for the backaction of the trapped impurity on the density  of the condensate near the potential edge, which is relevant in the small-$\gamma$ regime, due to the large compressibility of bose gas.
The impurity thus experiences an effective potential for small $x$:
\begin{equation*}
    V_\textrm{eff}(x) = \left\{ \begin{array}{c}
    \infty\qquad \qquad x\le 0 \\
    \gib n_0 \displaystyle{\frac{x^2}{2l_0^2}} \qquad x>0 .
\end{array}\right. 
\end{equation*}
The corresponding oscillator frequency $\omega$ follows from $V_\textrm{eff} = \frac{M}{2} \omega^2 x^2$. 
The ground-state energy of the impurity localized in $V_\textrm{eff}$ is thus
\begin{equation}
    E_\textrm{loc} \approx \frac{3 \omega}{2} = \frac{3}{2} \sqrt{\frac{\gib n_0}{M l_0^2}}
\end{equation}
 which scales only with the square root of $\gib$. Thus for a small impurity-boson interaction no bound state exists. A localized solution emerges only above a critical value 
 \begin{equation}
     \frac{\gib m}{n_0 }\Bigl\vert_\textrm{crit} = \frac{9}{4 \eta^2} \frac{m}{M} \frac{1}{n_0^2 \xi^2} =\frac{9}{2 \eta^2} \frac{m}{M} \gamma.\label{eq:critical-weak}
 \end{equation}
In \figref{fig:localization-weak}a we have plotted the critical transition line for $\eta=1$ (solid line). One recognizes that the backaction of the impurity to the condensate needs to be taken into account, which leads to an increased width of the density minimum of the condensate on the left edge, see \figref{fig:localization-weak}c. 

One can obtain an approximation for the factor $\eta$ in \eqref{eq:critical-weak} from a variational coherent-state ansatz considering a half-infinite system with only one edge at $x=0$.
To this end we assume a factorized ground state $\vert \psi_\textrm{gs}\rangle = \vert \phi\rangle\vert \phi_I\rangle$, with
\begin{equation}
    \phi(x) = \sqrt{n_0}\tanh\left(\alpha x\right).
\end{equation}
describing the coherent amplitude of the condensate.  $\vert \phi_I\rangle $ is then the ground state of the impurity in the resulting
effective P\"oschl-Teller potential, which can be calculated analytically \cite{nieto1978exact}. This approach has already been used to determine bound states of the impurity in a box potential using the ansatz $\alpha=1/(\sqrt{2}\xi)$ \cite{Petkovic2020}. Here we instead determine the characteristic length scale $\alpha$ by minimizing the total energy. This gives the implicit equation for 
$\eta = 1/(\sqrt{2} \xi \alpha)$ as function of the mass ratio $M/m$, the Tonks parameter $\gamma$ and  $\gamma_\textrm{IB}= \gib m/n_0$
\begin{align}
        0=&1-\eta^2  - \frac{15}{2}\frac{m}{M}\frac{\gamma^{1/2}}{\eta} + \frac{9}{2}\frac{m}{M}
        \frac{4\gamma_\textrm{IB}\frac{M}{m}\eta^2 +\gamma}{\eta \sqrt{8\gamma_\textrm{IB}\frac{M}{m}\eta^2+\gamma}}.      
       \label{eq:variationsparameter}
\end{align}
The solution of this equation is shown in \figref{fig:localization-weak} as a dashed line, showing very good agreement.

%%%%%%%%%%%%%%%%%%%%%%%%%%%%%%%%%%%%%%%%%%%%%%
%%%%%%%%%%%%%%%%%%%%%%%%%%%%%%%%%%%%%%%%%%%%%

In the Tonks gas limit $(\gamma \to \infty)$ the hardcore bosons are much less compressible but show Friedel oscillations in the density which also affect the impurity wavefunction in the localized phase. 
Hardcore bosons can be mapped to free fermions, which due to Pauli exclusion occupy all single-particle states in the trap up to the Fermi energy. Thus the density of the hardcore gas near one of the edges of the box is  given by
\begin{equation}
    n(x)\approx  \frac{2 k_F}{\pi}\left(1-\frac{\sin( 2k_F x)}{2 k_F x}\right),
\end{equation}
with $x$ being the distance from the edge and $k_F= \pi n_0/2$ is the Fermi momentum.
The characteristic length scale $l_0$ is here $(\sqrt{6}/\pi) n_0^{-1}$ which gives the following estimate for the critical $\gib$ of the localization-delocalization transition:
 \begin{equation}
     \frac{\gib m}{n_0 }\Bigl\vert_\textrm{crit} = \frac{9}{2} \frac{m}{M} \frac{\pi^2}{6}.\label{eq:critical-strong}
 \end{equation}
In \figref{fig:localization-strong}a.) we have also plotted this value for comparison and one recognizes reasonable good agreement. 
Due to the smaller compressibility of the Tonks gas, the localization transition is shifted to higher values of $\gib$.

%%%%%%%%%%%%%%%%%%%%%%%%%%%%%%%%%%%%%%%%%%%%%%
\begin{figure}[ht]
\centering
\includegraphics[width=0.8\columnwidth]{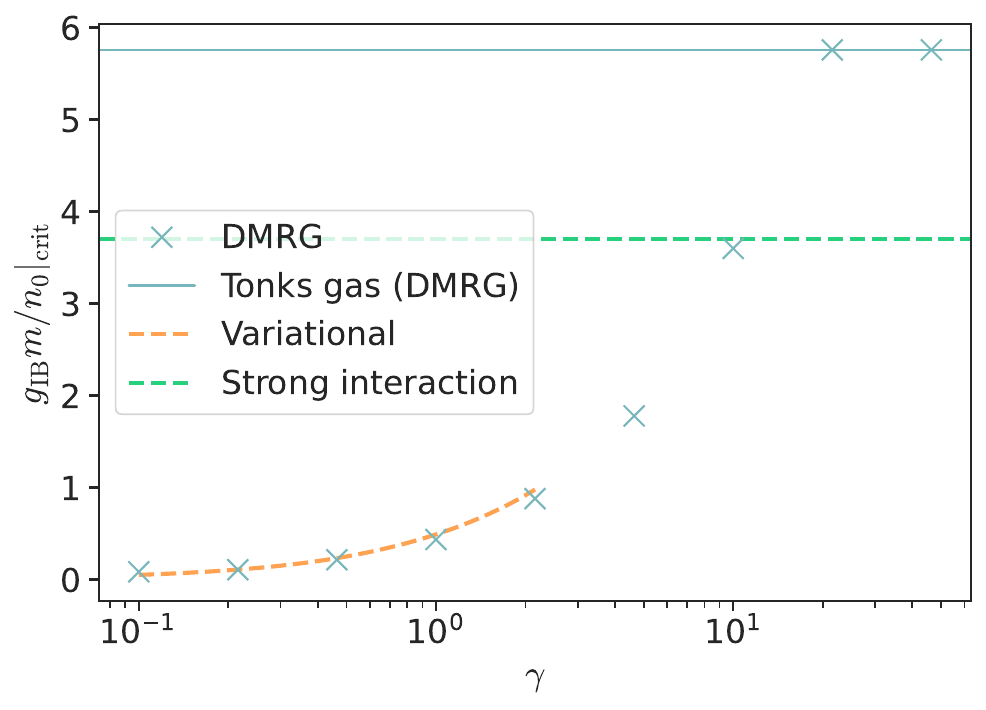}
\caption{Critical $\gib$ of the localization-delocalization transition for different Tonks parameters $\gamma$, a constant mass ratio $M/m=2$ and $N=40$ bosons calculated from DMRG simulations (crosses). Also shown is the critical $\gib$ derived from DMRG simulations of a Tonks gas $\gamma\rightarrow\infty$ (solid blue line). These values are compared to the predictions for a weak interacting Bose gas made by the variational ansatz (orange dashed line) and predictions for a Tonks gas \eqref{eq:critical-strong} (green dashed line). 
} \label{fig:localization-different_gamma} 
\end{figure}
%%%%%%%%%%%%%%%%%%%%%%%%%%%%%%%%%%%%%%%%%%%%%

In \figref{fig:localization-different_gamma} we compare $\gib|_\textrm{crit}$, obtained from DMRG simulations with the analytic predictions for small, \eqref{eq:critical-weak}, and large Tonks parameters $\gamma$, \eqref{eq:critical-strong}. Since the box potential is finite, we determine the transition point 
of the localization-delocalization transition by calculating at the uncertainty of the impurity position as function of $\gib$. At the critical value $\gib|_\textrm{crit}$ the position uncertainty is maximal. 
Analytical and numerical values align well up to a $\gamma\approx2.5$, above this value the variational ansatz fails to produce a real value of $\eta$.
For $\gamma>10$ the numerical values overshoot the analytic predictions from \eqref{eq:critical-strong}. This is because we approximated the potential produced by the Tonks gas as a harmonic potential.

%%%%%%%%%%%%%%%%%%%%%%%%%%%
\subsection{Absence of self-localization in a weakly interacting Bose gas}
%%%%%%%%%%%%%%%%%%%%%%%%%%%

\begin{figure}[h]
\centering
\includegraphics[width=\columnwidth]{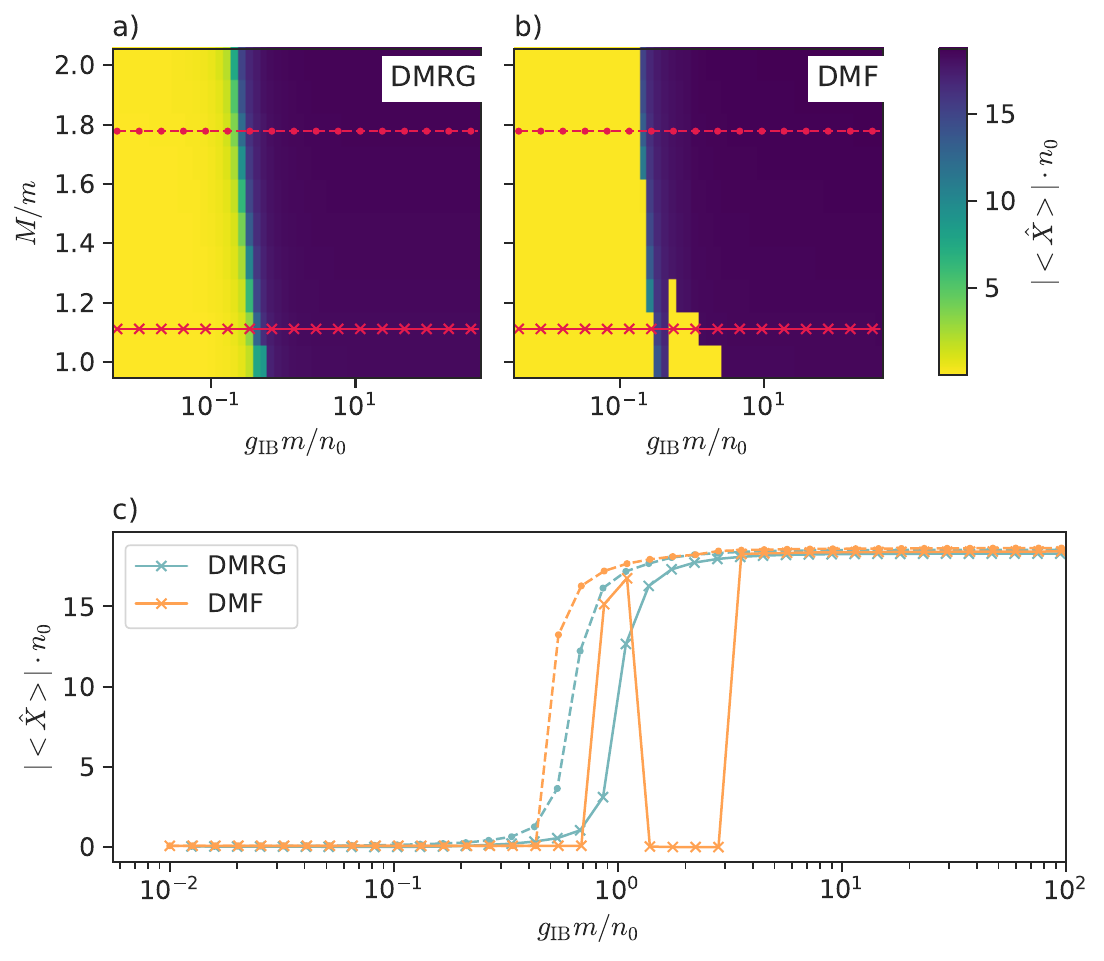}
\caption{Comparison between average impurity position $\expval{\hat{X}}$ from DMRG and DMF calculations for $\gamma=0.4$ and $N=40$. {a)} Colorplot of 
$\expval{\hat{X}}$ as function of 
  ${M}/{m}$ and $\gib$ obtained by DMRG. Red lines show cuts which are plotted in c). {b)} The same obtained from DMF.
 {c)} Cuts in a) for two mass ratios,  
solid and dashed line correspond to each other. One clearly recognizes the absence of a self-trapped solution in the DMRG simulations.} \label{fig:comparison-DMRG-SSM-Mean-position} 
\end{figure}

\begin{figure}[h!]
\centering
\includegraphics[width=\columnwidth]{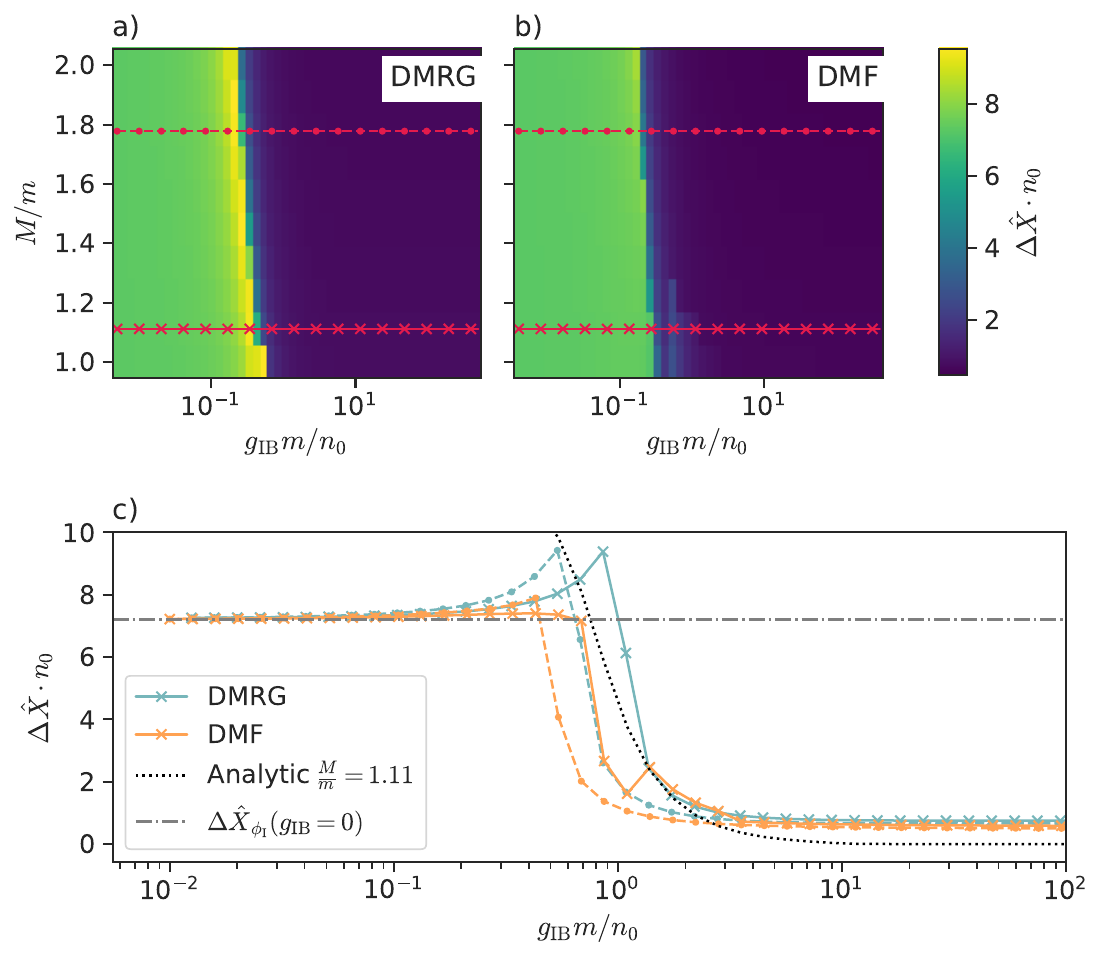}
\caption{Comparison between spatial variance of impurity 
$\langle\Delta \hat{X}\rangle$ 
from DMRG and DMF  for $\gamma=0.4$ and $N=40$. {a)} Colorplot of variance from DMRG simulations as function of ${M}/{m}$ and  $\gib$. Red lines show cuts which are plotted in c). {b)} 
The same obtained from DMF.
{c)} Cuts in a) for two mass ratios,  
solid and dashed line correspond to each other.
The grey dashed-dotted line shows the spatial width of the impurity expected for $\gib=0$.
%, which is given by the ground-state density distribution of a particle in a box $\cos^2{\left(x\pi/L\right)}$. 
The black dotted line shows the self-localization length $\lambda$ predicted in DMF for $\frac{M}{m}=1.11$. 
} \label{fig:comparison-DMRG-SSM-Variation} 
\end{figure}

We now compare the results from DMRG simulations with solutions of the DMF equations \eqref{eq:GPE-SE}. In \figref{fig:comparison-DMRG-SSM-Mean-position}
we show the average position of the impurity as function of the mass ratio $M/m$ and the impurity-boson interaction $\gib$ from DMRG (a) and DMF (b) simulations for $\gamma=0.4$, i.e. for weak boson-boson interactions. Figure (c) shows a cut for two mass ratios. Since the Tonks parameter is still small, we expect the mean-field simulations to provide good results. And indeed
one recognizes that the localization-delocalization transition to the edge of the trap is reasonably well captured by the DMF approach. However, for small mass ratios $\frac{M}{m}<1.3$ there is an area of coupling strengths around $\gib m/n_0 \approx0.5$ where there is a self-localisation of the impurity in the middle of the box. Importantly the width of the impurity distribution in this region is much smaller than the box size, as can be seen
in \figref{fig:comparison-DMRG-SSM-Variation}, where we compared the width of the impurity distribution from DMRG and DMF simulations.
%%%%%%%%%%%%%%%%%%%%%%%%%%%%%%%%%%%%%%%%%%%%%%
%%%%%%%%%%%%%%%%%%%%%%%%%%%%%%%%%%%%%%%%%%%%%

%%%%%%%%%%%%%%%%%%%%%%%%%%%%%%%%%%%%%%%%%%%%%%
%%%%%%%%%%%%%%%%%%%%%%%%%%%%%%%%%%%%%%%%%%%%%

%%%%%%%%%%%%%%%%%%%%%%%%%%%%%%%%%%%%%%%%%%%%%%
\begin{figure}[ht]
\centering
\includegraphics[width=\columnwidth]{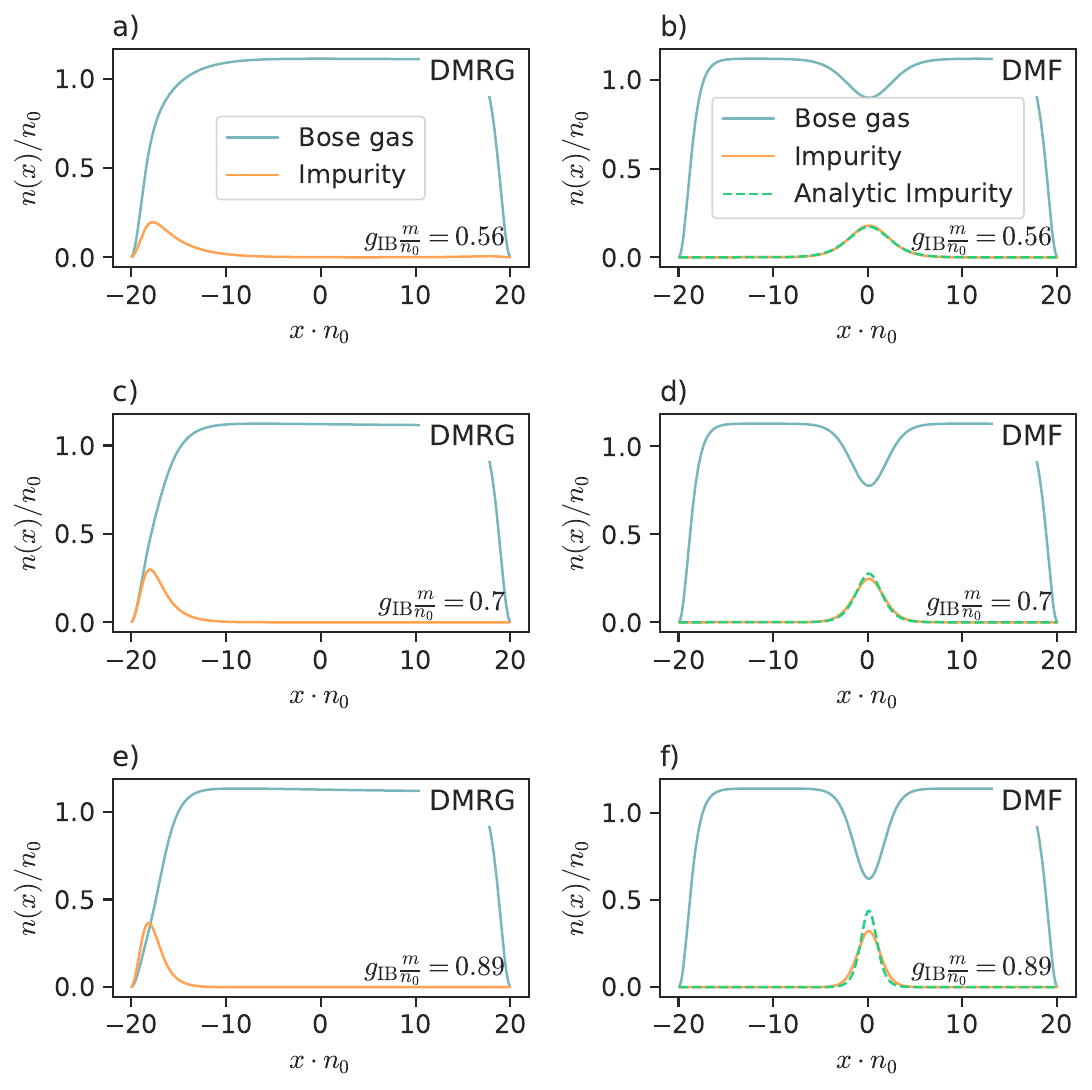}
\caption{\textit{Left:} Density distribution of Bose gas and impurity for different $\gib$ (given in the lower right) from DMRG simulations.
\textit{Right:} The same from mean-field calculations compared to the analytic prediction for a self-trapped polaron, \eqref{eq:analytic}. Parameters of the bose gas are $\gamma=0.4$, $M/m=1$ and $N=40$.} \label{fig:comparison_density_DMRG_SSM_analytik} 
\end{figure}
%%%%%%%%%%%%%%%%%%%%%%%%%%%%%%%%%%%%%%%%%%%%%

In \figref{fig:comparison_density_DMRG_SSM_analytik} we have plotted the density distributions of bosons and impurity in the box potential for a mass ratio $M/m=1$ for different values of $\gib$ inside the region of self trapping predicted by DMF.
Also shown is the analytic prediction from \cite{Bruderer-EPL2008} for the 
impurity probability density in the self-trapping regime
\begin{equation}
    \vert \phi_\mathrm{I}(x)\vert^2_\textrm{TF} = \frac{1}{2\lambda}\sech^2\left(\frac{x}{\lambda}\right) \label{eq:analytic}
\end{equation}
with $\lambda=\sqrt{2/\gamma^3}(g/\gib)^2(m/M)n_0^{-1}$ being the localization length,
which holds in the Thomas Fermi limit, i.e. for $\lambda \gg n_0^{-1}$.

In summary the self-localisation is an artifact of the decoupling approximation and is not present in the full quantum approach. This is because
boson-impurity correlations, which are neglected in DMF, suppress the self-localization \cite{zschetzsche2024suppression}. Different from \cite{zschetzsche2024suppression} self localization does not occur in the full quantum simulations even for large impurity-boson couplings.

%%%%%%%%%%%%%%%%%%%%%%%%%%%
%%%%%%%%%%%%%%%%%%%%%%%%%%%
\section{Heavy polaron in a bose gas with arbitrary boson-boson interactions}
\label{sec:polaron}
%%%%%%%%%%%%%%%%%%%%%%%%%%%
%%%%%%%%%%%%%%%%%%%%%%%%%%%

In \cite{Jager2020,Will2021} we have shown that key properties of heavy polarons in 1D Bose gases can be rather accurately be described by a mean-field theory in a LLP frame that takes into account the backaction of the impurity to the condensate.  Quantum fluctuations have to be included only in a linearized  approximation. 
These studies were however limited to weak interactions of the bosons, characterized by Tonks parameter $\gamma < 1$. In the following we show that the opposite case of large Tonks parameters $\gamma\gg 1$ can be accurately treated as well by a perturbation expansion of a complementary mean-field description in the Tonks gas limit.

A characteristic quantity of the polaron is the energy $E_\mathrm{P}$  needed to immerse an impurity into the Bose gas $E_\mathrm{P}=E(\gib\neq0)-E(\gib=0)$.
Here $E(\gib=0)$ is the ground state energy of the system without Bose-impurity-interaction. In the following we will compare analytic and semi-analytic predictions from linearized fluctuation expansions around mean field approaches in the weak- ($\gamma <1$) and strong-interaction cases ($\gamma\gg 1$) with exact numerical predictions. To this end we will employ DMRG simulations, which agree with previous exact results obtained by Quantum Monte Carlo simulations in Ref.\cite{Parisi2017a}.

%%%%%%%%%%%%%%%%%%%%%%%%%%%
\subsection{Polaron energy in a weakly interacting Bose gas}
%%%%%%%%%%%%%%%%%%%%%%%%%%%

For a heavy impurity and a weakly interacting Bose gas, $\gamma \ll 1$, with periodic boundary conditions, a mean-field approximation to the polaron Hamiltonian after a Lee-Low-Pines transformation gives a rather accurate prediction of the polaron energy 
\cite{Volosniev2017,Panochko2019,Jager2020}:
\begin{align}
    E_P=\frac{4}{3} n\overline{c} \left[1+ \frac{3}{2}\chi+\chi^3-\left(1+\chi^2\right)^{3/2}\right],\label{eq:E-polaron}
\end{align}
where the dimensionless parameter $\chi=\gib/(2\sqrt{2}gn\overline{\xi})$ characterizes the impurity-boson interaction strength. If $|\chi|\gtrapprox1$ the condensate undergoes substantial deformation.
$\overline{\xi} =\xi \sqrt{m/m_r} $ and $\overline{c} = \sqrt{m/m_r}c$ are the rescaled healing length and speed of sound respectively.
 $n$ is the boson density without impurity and $m_r =(Mm)/(M+m)$ is the reduced mass.
%
%
%\begin{equation}
%    E_p =\gib n_0 \left(\frac{y-1}{y+1}\right)^2 + \frac{8}{3}n_0 \overline{c} \left(\frac{3y+1}{(y+1)^3}\right),\label{eq:E-polaron}
%\end{equation}
%
% where $y= \sqrt{1 +8 (g n_0 \overline{\xi}/\gib)^2} + \sqrt{8 (g n_0^2 \overline{\xi}/\gib)^2}$ with $\overline{\xi} =\xi \sqrt{m/m_r} $ and $\overline{c} = \sqrt{m/m_r}c$ being the rescaled healing length and speed of sound respectively. $n_0$ is the boson density without impurity and $m_r =(Mm)/(M+m)$ is the reduced mass.

%%%%%%%%%%%%%%%%%%%%%%%%%%%
\subsection{Polaron energy in a strongly-interacting Bose gas}
%%%%%%%%%%%%%%%%%%%%%%%%%%%

In the following we will discuss the case of a strongly interacting Bose gas, $\gamma \gg 1$.
To this end we consider an impurity whose mass is much larger then that of the bosons ($\frac{M}{m}\gg1$) such that  the kinetic energy of the impurity can be neglected. 
The resulting effective Hamiltonian then reduces to a interacting boson problem in an external $\delta$ potential 

\begin{align}
    \hat{H}=\int\dx\, \hat{\phi}^\dagger(x)\left(-\frac{\partial^2_x}{2m}+\frac{g}{2}\hat{\phi}^\dagger(x)\hat{\phi}(x)+\gib\delta\left(x\right)\right)\hat{\phi}(x).\label{eq:H_Bose_impurity_infinite_mass}
\end{align}

\begin{figure}[ht]
\centering
\includegraphics[width=\columnwidth]{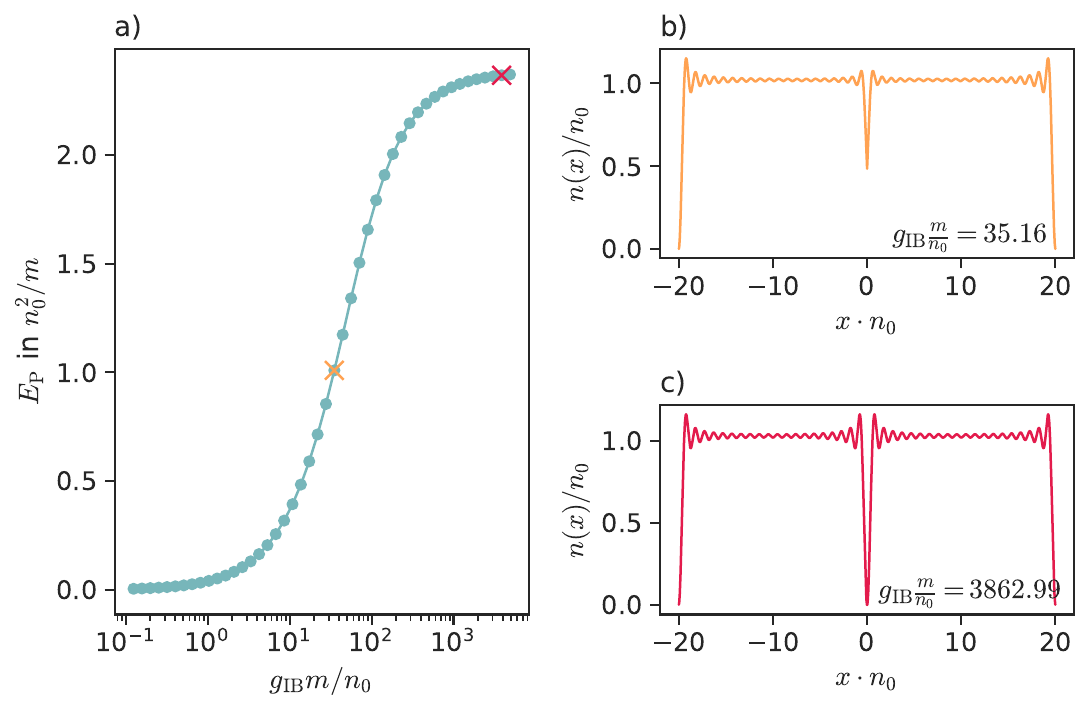}
\caption{Polaron in strongly interacting 1D Bose gas with $\gamma=20$ and $N=80$ particles in a finite box and $M=\infty$: {a)} polaron energy as function of impurity-boson interaction, {b)}, {c)} wavefunction for two different values of $\gib$ (written in the lower right) in a box potential and impurity fixed at $x=0$. 
} \label{fig:energy-2} 
\end{figure}

Figure \ref{fig:energy-2} shows the polaron energy in a strongly ($\gamma=20)$ interacting, trapped 1D Bose gas. The boson density shows Friedel like oscillations, see also \cite{Parisi2017a}, and the  
characteristic size of the condensate depletion around the fixed position of the impurity is given by  the wavelength of these oscillations.

%%%%%%%%%%%%%%%%%%%%%%%%%%%%%%%%%%%%%%%%%%%%%%
%%%%%%%%%%%%%%%%%%%%%%%%%%%%%%%%%%%%%%%%%%%%%

%%%%%%%%%%%%%%%%%%%%%%%%%%%
\subsubsection{Tonks limit $\gamma\to\infty$}
%%%%%%%%%%%%%%%%%%%%%%%%%%%

In the limit $\gamma\to\infty$ the 1D interacting Bose gas maps to free fermions. Here we can 
calculate the single particle solutions of free fermions with a infinitely heavy impurity at $x=0$ for open boundary conditions (OBCs). This system is described by the Hamiltonian
\begin{align}
    \hat{H}_F^0=\int_{-L/2}^{L/2} \!\!\!\! \dx\, \, \hat{\psi}^\dagger(x)\left(-\frac{\partial^2_x}{2m}+\gib\delta\left(x\right)\right)\hat{\psi}(x),\label{eq:H_free_Fermion_impurity}
\end{align}
with $\{\hat \Psi(x),\hat\Psi^\dagger(y)\}=\delta(x-y)$.
This means we need to solve the problem of non-interacting fermions trapped in a infinitely deep box potential (OBC's) with a delta potential in the middle.

The ground state $\ket{\psi}$ and the ground state energy $E$ can be written as
\begin{align}
    \ket{\psi} =& \int \!\!\dx_1 \dots \int\!\! \dx_N\,  \varphi_1\left(x_1\right) \dots \varphi_N(x_N)\times \nonumber\\
    &\qquad\times \hat{\psi}^\dagger(x_1)\dots \hat{\psi}^\dagger(x_N)\ket{0},\label{eq:fermi_state}\\
   & E=\sum_{l=1}^N\frac{k_l^2}{2m}\label{eq:fermi_energy},
\end{align}
where the single particle solutions are given by 
\begin{align}
    \varphi_l(x)=a_l\mathrm{sgn}(x)^l\sin\left(k_l\left(|x|-\frac{L}{2}\right)\right).
\end{align}
$l$ is an integer and the wave numbers $k_l$ have to be numerically calculated from the boundary condition at the delta potential
\begin{align}
    0 = \frac{k_l}{2m}\left(1+(-1)^l\right) + \gib\tan{\left(k_l\frac{L}{2}\right)}.\label{eq:k_values}
\end{align}
This then gives for the polaron energy, see \cite{Parisi2017a}
\begin{eqnarray}
     E_p &=& \frac{n_0^2 \pi^2}{2 m} \varepsilon(\gib),\nonumber\\
     \varepsilon(\gib) &=& \frac{1}{\pi}\left[\left(1+ \frac{\eta^2}{\pi^2}\right)\arctan\frac{\eta}{\pi}+\frac{\eta}{\pi}- \frac{\eta^2}{2\pi}\right],
\end{eqnarray}
where $\eta = \gib m/n_0 >0$.

%%%%%%%%%%%%%%%%%%%%%%%%%%%%%%%%%%%%%%%%%%%%%
\subsubsection{Analytic approximation of the polaron energy for $1\ll \gamma < \infty$.}
%%%%%%%%%%%%%%%%%%%%%%%%%%%%%%%%%%%%%%%%%%%%%

For large but finite values of $\gamma$ one can derive semi-analytic expressions for the polaron energy.
As shown by Girardeau \cite{Girardeau1960}, 
bosons with $s$-wave contact interactions are dual to spin-polarized fermions with $p$-wave interactions and both can be 
mapped onto each other by the well-known boson-fermion mapping \cite{Girardeau1960,cheon1999fermion}:
The bosonic Hamiltonian \eqref{eq:H_Bose_impurity_infinite_mass} can be written as a fermionic Hamiltonian
\begin{equation}
    \hat{H}_F = \hat{H}_F^0 + \hat{H}_1
\end{equation}
with
\begin{align}
        \hat{H}_1 =  
       \label{eq:H_F}
         -\frac{g_\mathrm{F}}{2}\int\!\!\!\int \dx\, \dy\,\, \hat{\psi}^\dagger(x)\hat{\psi}^\dagger(y)\overset{\leftarrow}{\frac{\partial}{\partial z}}\delta(z)\overset{\rightarrow}{\frac{\partial}{\partial z}}\hat{\psi}(y)\hat{\psi}(x),
\end{align}
where $z = x-y$ and $\hat{\psi}$
are fermionic field operators. The arrows above the derivatives mean that in the case of the arrow to the left, the derivative has to applied to the function on the left of it and for the arrow to the right on the function to the right. The $p$-wave interaction strength $g_\mathrm{F}$ is then related to the bosonic interaction strength $g$ via
\begin{align}
    g_\mathrm{F}=-\frac{4}{g}.
\end{align}
This means a strongly $s$-wave interacting Bose gas maps to a weakly $p$-wave interacting Fermi gas and vice versa. 

The boson-fermion mapping provides an elegant way to find approximate analytic expressions for the polaron energy in a strongly interacting Bose gas with $\gamma \gg 1$ perturbatively in $\vert g_\mathrm{F}\vert \sim 1/\gamma$.

%%%%%%%%%%%%%%%%%%%%%%%%%%%%%%%%%%%%%%%%%%%%%%
\begin{figure}[ht]
\centering
\includegraphics[width=0.9\columnwidth]{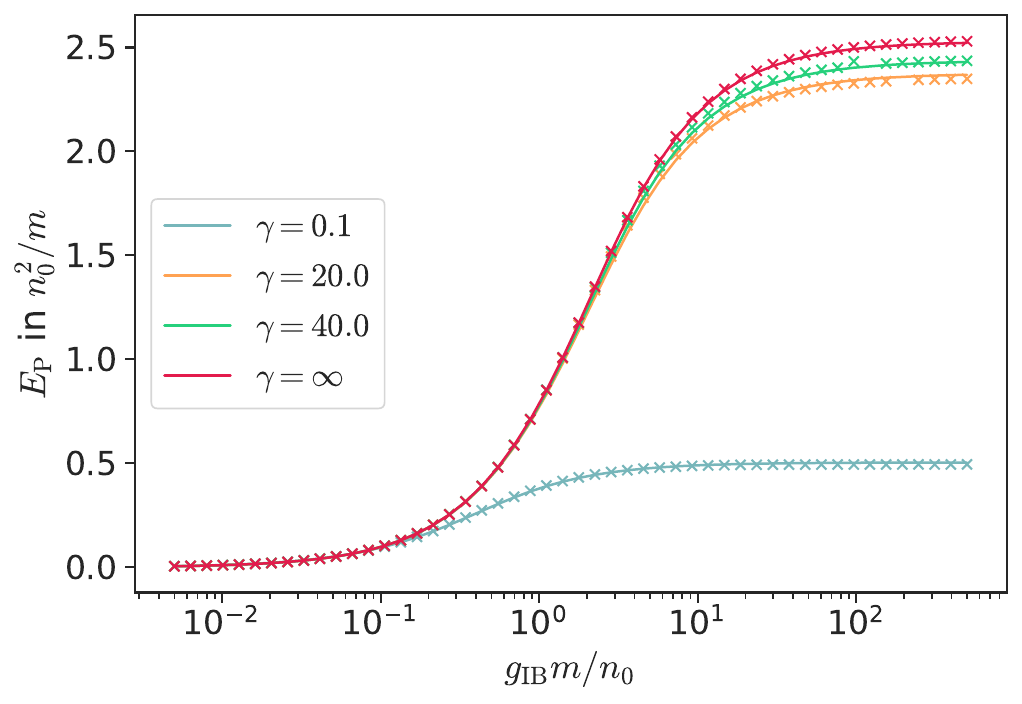}
\caption{Comparison of perturbative analytic approximation to polaron energy (solid lines) with DMRG results (crosses) for a strongly interacting 1D Bose gas with $\gamma=20, 40, \infty$ and $N=80$ particles respectively and $M\to\infty$. In light blue are DMRG results (crosses) compared to mean-field results (solid line) \cite{Jager2020,Will2021} for a weakly interacting Bose gas with $\gamma=0.1$. 
} \label{fig:energy-comparison} 
\end{figure}
%%%%%%%%%%%%%%%%%%%%%%%%%%%%%%%%%%%%%%%%%%%%%

The second line in \eqref{eq:H_F} can now be treated as perturbation, i.e. $\hat H_F= \hat H_F^0 + \hat H_1$ and the first-order energy correction $\Delta E_1$ reads
\begin{align}
    \Delta E_1=\bra{\psi}\hat{H}_1\ket{\psi}\\
    \begin{split}
        =\frac{g_\mathrm{F}}{2}\sum_{l,n\neq l}^N\frac{a_l^2a_n^2k_l}{4k_n}\biggl[&\bigl(k_n-k_n\cos{\left(k_n L\right)}\bigr)\sin{\left(k_l L\right)}\\
        &+k_l\bigl(k_n L-\sin{\left(k_n L\right)}\bigr)\biggr].
    \end{split}
\end{align}
\figref{fig:energy-comparison} shows a comparison of the polaron energies as function of $\gib$ from first order perturbation in $1/\gamma$, as well as from \eqref{eq:E-polaron} for small values of $\gamma$, with numerical DMRG results. 
In both cases the polaron energy saturates for increasing $\gib$, which 
happens when the density of the Bose gas at the position of the impurity approaches zero.
One recognizes rather good agreement between analytic approximations and exact simulations both for small and large values of $\gamma$  even down to $\gamma=20$ and for all values of $\gib$. Note that the analytic approximation for $\gamma=\infty$ is that from Ref.\cite{Parisi2017a}.

Although we have considered here only the case of an impurity with an infinite mass and large values of $\gamma \ge 10$ and $\gamma \le 0.1$ one recognizes that key properties of polarons formed by heavy impurities in a 1D Bose gas with arbitrary boson-boson and arbitrary boson-impurity interactions can be obtained rather accurately from a proper mean-field approach and potentially including lowest-order quantum corrections in Bogoliubov approximation.

%%%%%%%%%%%%%%%%%%%%%%%%%%%
%%%%%%%%%%%%%%%%%%%%%%%%%%%
\section{Polaron interaction in a strongly interacting Bose gas in the Born-Oppenheimer limit} \label{sec:bipolaron}
%%%%%%%%%%%%%%%%%%%%%%%%%%%
%%%%%%%%%%%%%%%%%%%%%%%%%%%

Interactions between particles mediated by a many-body environment play an important role in condensed-matter systems. Examples include the Ruderman-Kittel-Kasuya-Yodsia (RKKY) interaction of spins in a Fermi liquid \cite{Ruderman1954,Kasuya1956,Yosida1957} and Cooper pairing of electrons \cite{Cooper1956}. The mechanism responsible for these interactions is the same as what causes the formation of quasi-particles such as the polaron. 
Even if the  impurities don't interact directly with each other, as we will assume here, they do so by their coupling to the condensate. If the impurities get close to each other they expel the Bose gas between them and the surrounding gas pushes the impurities together. This causes an effective attractive interaction  which in itself can lead to the formation of bound states called bi-polaron states.
The understanding of bi-polarons is 
one of the key questions of many-body physics.
They are suspected to be key for high-temperature superconductivity \cite{Alexandrov1992,Mott1993,Alexandrov1994} and phenomena such as 
the electric conductivity of polymers \cite{Bredas1985,Glenis1993,Bussac1993,Fernandes2005,Zozoulenko2019} or organic magneto-resistance \cite{Bobbert2007}.

For a weakly interacting Bose gas $(\gamma\ll 1)$ a mean-field ansatz in the LLP frame can be used to obtain semi-analytic expressions of the interaction potential at short distances, which agree very well with quantum Monte-Carlo simulations
\cite{Will2021}. The mean field approach does not describe the Casimir-like contributions arizing from the exchange of virtual phonons \cite{recati2005casimir,fuchs2007oscillating,Schecter2016,Reichert2019,petkovic2022mediated,schecter2014phonon}, which give however only important corrections in the tails of the interaction potential.

In this section we investigate the interaction between Bose polarons in a strongly-interacting 1D quasi condensate. In the Tonks limit $\gamma \to\infty$, where the interacting bosons can be mapped to free fermions, an analytic approximation to the interaction potential
has been obtained in Ref.\cite{recati2005casimir,fuchs2007oscillating} using a low-energy Luttinger-liquid approximation
\begin{equation}
    V(r) \approx \frac{v_F}{2\pi r} \textrm{Re}\, \textrm{Li}_2\left(-\frac{\gib^2 e^{2\pi  i n_0 r}}{\bigl(v_F+ i \gib\bigr)^2}\right),\quad r\gg n^{-1}\label{eq:LL-potential}
\end{equation}
with $v_F = \pi n_0/m$ being the Fermi velocity, and $\textrm{Li}_2$ the dilogarithmic function. One notices an oscillatory behavior with the frequency being that of Friedel oscillations. The low-energy approximation gives an accurate description of the large distance behavior of the interaction potential including the Casimir contributions, but fails at short distances, which are however important for the formation of bi-polaron bound states. Thus we here determine the interaction potential both numerically by DMRG simulations and analytically in the Tonks-gas limit for all distances $r$.
Since DMRG is not well suited for periodic boundary conditions, we again assume a confinement of the Bose gas to a box potential. As long as the 
distance of the impurities from the edges of the box  
is much larger than the healing length or the Fermi wavelength, boundary effects can be neglected. 
For the same reason our simulations, although in principle suitable, do not allow to accurately extract the far tails of the interaction potential, dominated by virtual phonon exchange. As discussed in Appendix B, the DMRG simulations reproduce the semi-analytical results from Ref.\cite{Will2021} obtained in mean-field theory for small values of $\gamma$.  

In the following we calculate the polaron interaction potential in Born-Oppenheimer approximation where $\frac{M}{m}\gg1$. Here  
the impurities do not posses kinetic energy and are localized in space. By varying their distance $r$ one can determine the polaron interaction potential $V(r)$ from 
\begin{equation}
    V(r)=E_{lr}(r)-E_l(r)-E_r(r)+E_0.\label{eq:bipolaron_potential}
\end{equation}
Here $E_{lr}(r)$ is the ground state energy with both impurities present, $E_l$ ($E_r$) the ground state energy with only the left (right) impurity, and $E_0$ is the ground state energy of the Bose gas without any impurities. In the case of two impurities and for $\frac{M}{m}\gg1$ the Hamiltonian
becomes
\begin{align}
\begin{split}
    \hat{H}_\mathrm{B}=&\int \dx\, \hat{\phi}^\dagger(x)\left(-\frac{\partial^2_x}{2m}+\frac{g}{2}\hat{\phi}^\dagger(x)\hat{\phi}(x)\right)\hat{\phi}(x)\\
    &+\hat{\phi}^\dagger(x)\Bigl\{\gib\left[\delta\left(x+\frac{r}{2}\right)+\delta\left(x-\frac{r}{2}\right)\right]\Bigr\}\hat{\phi}(x),
\end{split}\label{eq:H_Bose_bipolaron_infinite_mass}
\end{align}
where the two impurities are described by the delta potentials at $\frac{r}{2}$ and $-\frac{r}{2}$. 

%%%%%%%%%%%%%%%%%%%%%%%%%%%%%%%%%%%%%%%%%%%%%%
\begin{figure}[ht]
\centering
\includegraphics[width=\columnwidth]{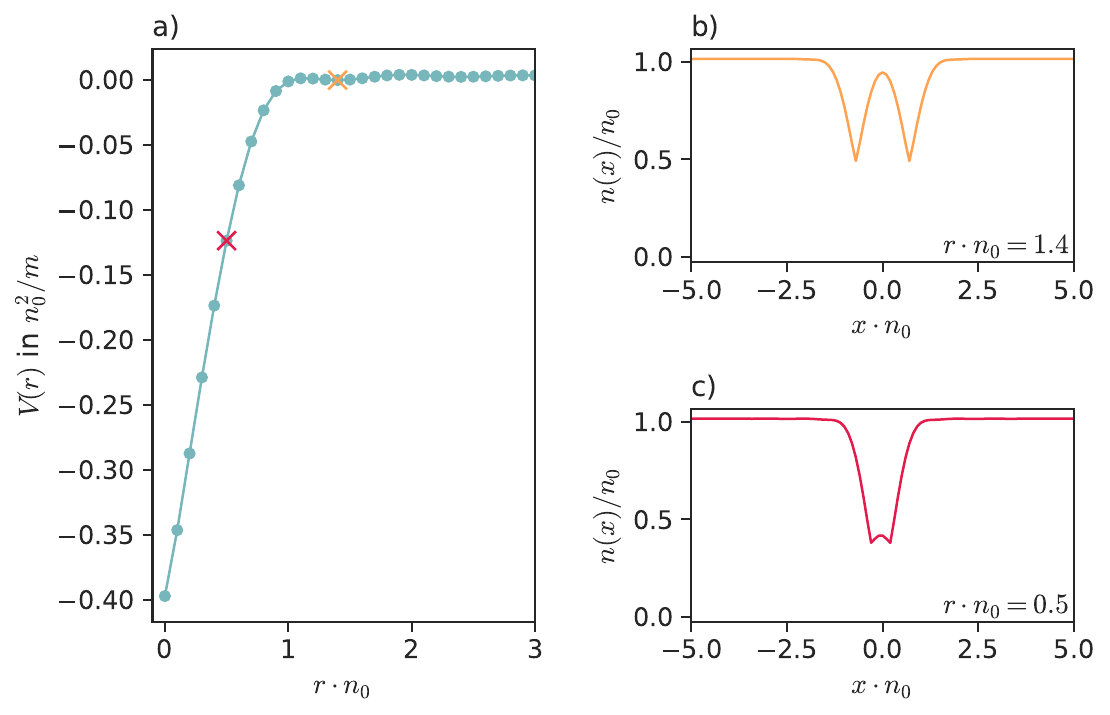}
\caption{\textit{Left:} Interaction potential of heavy impurities in a bose gas with $\gamma=4$ and $N=100$ particles and bose-impurity interaction strength $\gib m/n_0=1.08$. \textit{Right:} Density of Bose gas at the two separations indicated in left picture as crosses in the corresponding color.
} \label{fig:bipolaron-weak} 
\end{figure}
%%%%%%%%%%%%%%%%%%%%%%%%%%%%%%%%%%%%%%%%%%%%% 

\figref{fig:bipolaron-weak} shows the short-distance behaviour of $V(r)$ for a medium-sized Tonks parameter $\gamma=4$ as well as density distributions of the bosons for two different separations $r$ between the impurities. One recognizes a linear potential for small distances similar to the results of \cite{Will2021} for a weakly interacting Bose gas. 
The effective potential becomes attractive as soon as the Bose gas between the polarons is substantially diminished. The pressure from the atoms to the left and to the right of the polaron pair then causes a constant force and thus a linear interaction potential.

%%%%%%%%%%%%%%%%%%%%%%%%%%%%%%%%%%%%%%%%%%%%%%
\begin{figure}[ht]
\centering
\includegraphics[width=\columnwidth]{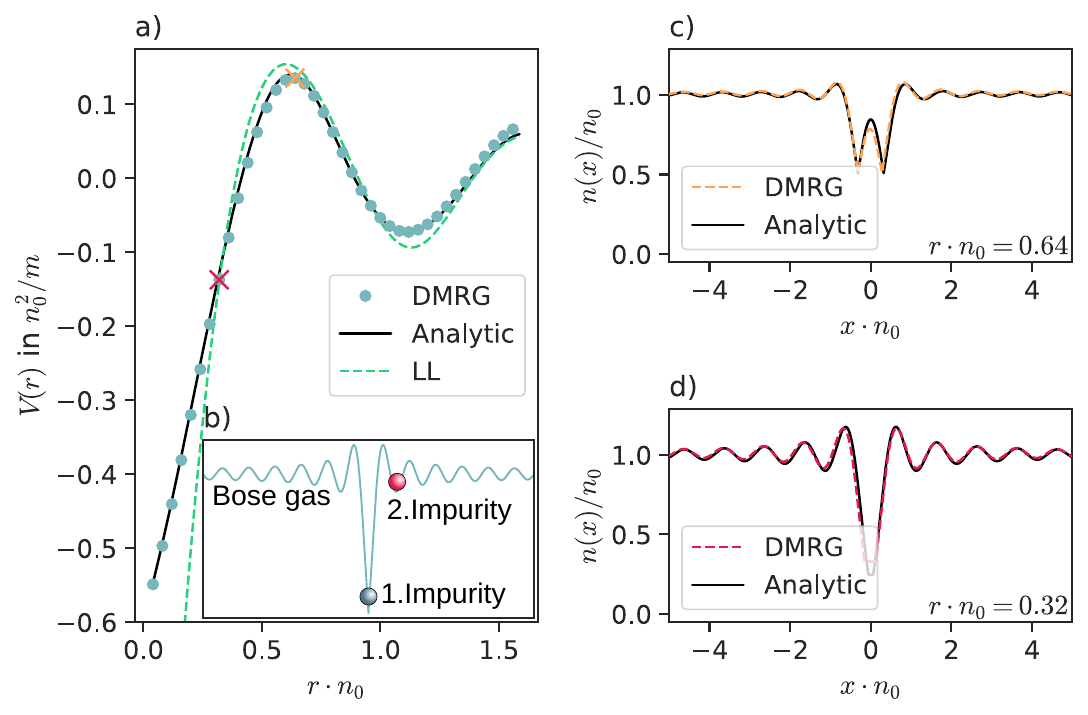}
\caption{\textit{Left:} Interaction potential of heavy impurities in the Tonks-gas limit $\gamma=\infty$, $N=80$ and $g_\textrm{IB}m/n_0=1.58$ The dashed line shows the Luttinger-liquid approximation from Ref.\cite{recati2005casimir}, \eqref{eq:LL-potential}. Insert illustrates origin of oscillatory potential. Also shown is the analytic prediction (black line) of the interaction potential obtained from mapping to free fermions in PBC. \textit{Right:} Density of Bose gas at the two separations indicated in left picture as crosses in the corresponding color.} \label{fig:bipolaron-strong} 
\end{figure}
%%%%%%%%%%%%%%%%%%%%%%%%%%%%%%%%%%%%%%%%%%%%%

For large values of $\gamma$ on the other hand oscillations appear in the polaron potential, as can be seen in \figref{fig:bipolaron-strong}. The frequency of these oscillations is that of the Friedel oscillations caused by the polarons themselves.   
As shown in \figref{fig:bipolaron-strong}b.), once a single impurity is introduced into the system it causes density oscillation and the second impurity  has to displace a smaller or a larger amount of bosons and therefore needs less or more energy to form a polaron. The resulting energy modulations reflect themselves in the polaron potential. Also shown is a comparison to the analytic low-energy approximation, \eqref{eq:LL-potential}, which diverges as $r\to 0$.

It can also be seen, that the gradient of the potential increases with the Tonks parameter.  This is the case since the force results from the surrounding Bose gas pushing against the polarons. For a stronger interacting Bose gas the resulting larger quantum pressure in the gas exerts a bigger force on the polarons. 

The interaction potential in the Tonks gas limit can also be obtained 
directly. To this end we have to calculate the energy of a free Fermi gas (Tonks gas) in a box with periodic boundary conditions in the presence of one or two $\delta$-potentials with strength $\gib$ at positions $x=\pm r/2$, which is an elementary quantum mechanics problem.
The interaction potential can then be obtained from the total energy $E_\mathrm{lr}$ of the Tonks gas in the presence of two $\delta$-potentials, the energy $E_1$ of a single $\delta$-potential, and the energy without impurities $E_0$:
\begin{align}
    V(r)=E_\mathrm{lr}-2E_1+E_0.
\end{align}
As can be seen in \figref{fig:bipolaron-strong} the bipolaron potential as well as the density distribution of the Tonks gas obtained in this way compare quite well with the DMRG simulations. These results are also in agreement with earlier findings in \cite{huber2019medium}.

%%%%%%%%%%%%%%%%%%%%%%%%%%%
%%%%%%%%%%%%%%%%%%%%%%%%%%%
\section{Summary}
%%%%%%%%%%%%%%%%%%%%%%%%%%%
%%%%%%%%%%%%%%%%%%%%%%%%%%%

In the present paper we studied the ground state of a single and two impurities in a one-dimensional Bose gas for arbitrary impurity-boson and boson-boson interactions, addressing: (i) the existence of self-localization of polarons, (ii) the accuracy of mean-field descriptions of polarons which take impurity-boson correlations into account, and (iii) the Born-Oppenheimer interaction potential between two polarons beyond a low-energy approximation.   
To fully account for quantum effects within the Bose gas, which are particularly important in the limit of large Tonks parameters $\gamma$, we performed numerical simulations  of a discretized effective lattice model using the density-matrix renormalization group. 

In a commonly used mean-field approach to the Bose polaron, condensate and impurity are described by a factorized impurity-boson c-number wavefunction, leading to a coupled Gross-Pitaevski -- Schr\"odinger equation. Such an approach predicts the existence of a self-trapped polaron for arbitrarily small impurity-boson couplings in a homogeneous 1D gas, where the center of mass of the impurity is localized in a distortion of the condensate created by the impurity. Such a decoupling mean-field (DMF) theory neglects spatial correlations between impurity and bosons. Recent findings
have shown that including these correlations in leading order prevents self-trapping for small
impurity-boson couplings \cite{zschetzsche2024suppression}. We here showed by comparison with exact DMRG results that the self-trapped solution is an artifact of the DMF approach and does not exist also for large impurity-boson interactions.

Mean-field approaches to the Bose polaron in a frame of relative coordinates between bosons and impurity, obtained by a Lie-Low-Pines transformation, amended by a linearized fluctuation analysis,
have been shown to provide accurate descriptions of Bose polarons for heavy impurities ($M/m\ge 3$) and weak boson-boson interactions $(\gamma\le 1)$
\cite{Volosniev2015,Volosniev2017,dehkharghani2015quantum,Jager2020,Will2021}. 
We here showed that the same is true for strongly interacting bosons. To this end we calculated the polaron energy and derived analytical approximations for large but finite Tonks parameters $1\ll\gamma <\infty$ and arbitrary boson-impurity couplings by using the mapping to weakly interacting fermions.

Finally we numerically calculated the short-distance interaction potential between two impurities in Born-Oppenheimer approximation for arbitrarily strong boson-boson interactions. For small Tonks parameter $\gamma \le 1$ we verified the results of \cite{Will2021} were a linear short-distance behavior was predicted. In the strong interaction limit $\gamma\gg 1$ we found oscillatory modulations in the potential in agreement with low-energy approximations \cite{recati2005casimir,fuchs2007oscillating} and extending them to short distances, relevant for bi-polaron bound states.

\emph{Note}: After submission of our manuscript a work by Gomez-Lozada \textit{et al.} appeared studying polarons in 1D lattices of interacting bosons \cite{gomez2024bose}, also showing (among other things) that correlations prevent phase separation and self-trapping.

%%%%%%%%%%%%%%%%%%%%%%%%%%%
\subsection*{Acknowledgement}
%%%%%%%%%%%%%%%%%%%%%%%%%%%

The authors gratefully acknowledge financial support by the DFG through SFB/TR 185, Project No.277625399. E.V.M. furthermore acknowledges support by Fundació Agustí Pedro Pons.

%%%%%%%%%%%%%%%%%%%%%%%%%%%
%%%%%%%%%%%%%%%%%%%%%%%%%%%
\section*{Appendix} 
%%%%%%%%%%%%%%%%%%%%%%%%%%%
%%%%%%%%%%%%%%%%%%%%%%%%%%%

%%%%%%%%%%%%%%%%%%%%%%%%%%%%%%%%%%%%%%%%%%%%%%%%%%%%%%%%%%%%%%%%%%%%%
\subsection{DMRG simulation of quantum impurity in a 1D Bose gas}
%%%%%%%%%%%%%%%%%%%%%%%%%%%%%%%%%%%%%%%%%%%%%%%%%%%%%%%%%%%%%%%%%%%%%

In order to apply DMRG for the continuum model 
let us first look at the Hamiltonian where both bosons and the impurity are described in second quantization:
\begin{align}
    \begin{split}
        \hat{H}&=\int\dx\,\hat{\phi}^\dagger(x)\left[-\frac{\partial^2_x}{2m}+\frac{g}{2}\hat{\phi}^\dagger(x)\hat{\phi}(x)\right]\hat{\phi}(x)\\
        &+\int\dx\,\hat{\phi}_\mathrm{I}^\dagger(x)\left[-\frac{\partial^2_x}{2M}+\gib\hat{\phi}^\dagger(x)\hat{\phi}(x)\right]\hat{\phi}_\mathrm{I}(x),
    \end{split}\label{eq:konti_ham_disc}
\end{align}
where $\hat{\phi}_\mathrm{I}$ is the impurity field operator.

\begin{figure}[ht]
\centering
\includegraphics[width=\columnwidth]{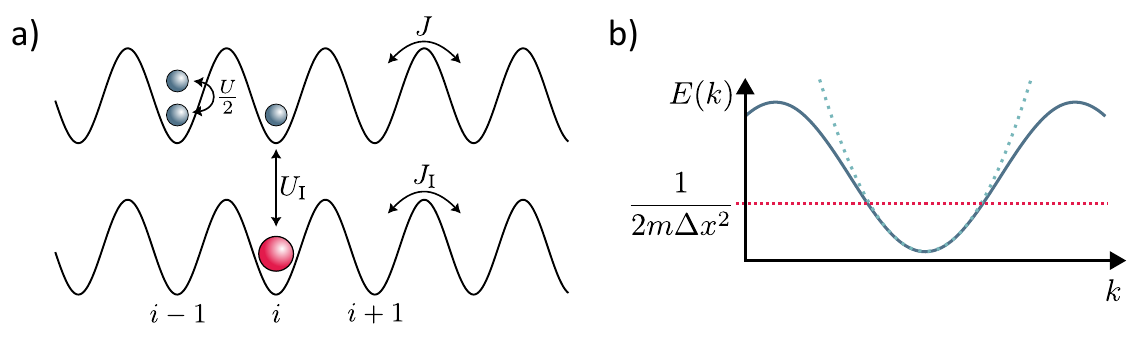}
\caption{{a)} Illustration of terms in Hamiltonian \eqref{eq:discretized_ham_hp}. The blue circles depict bosons which interact with strength $\frac{U}{2}$ and can hop with amplitude $J$. The impurity is shown as a red circle. It interacts with the bosons with  strength $U_\mathrm{I}$ and hops between lattice sites with  amplitude $J_\mathrm{I}$. {b)} Comparison between the sinusoidal dispersion relation of particles in a lattice to that of free particles, with the maximal energy to which both remain comparable.
} \label{fig:sketch_discret_ham} 
\end{figure}

Discretization of the $x$-coordinate into a 1D lattice with lattice spacing $\Delta x$, the field operators can be replaced by creation and annihilation operators at position $x_i = i\Delta x$
\begin{align}
    \hat{\phi}(x_i)\rightarrow\frac{1}{\sqrt{\Delta x}}\hat{a}_i\qquad
    \hat{\phi}_\mathrm{I}(x_i)\rightarrow\frac{1}{\sqrt{\Delta x}}\hat{b}_i.
\end{align}
The second order derivative in \eqref{eq:konti_ham_disc} can then be written as
\begin{align}
    \frac{\partial^2}{\partial x^2} \hat{\phi}(x) &\approx \frac{\hat{a}_{i-1}-2\hat{a}_{i}+\hat{a}_{i+1}}{\Delta x},\label{eq:discret_second_deriv}
\end{align}
and we set the hoppings which go from and to site $0$ and $L+1$ to $0$. This is justified because our average system sizes exceed $400$ lattice sites were the energy contributions from these hopping terms are small compared to the overall energy.
By inserting these transformations into \eqref{eq:konti_ham_disc} one arrives at a tight-binding lattice Hamiltonian
\begin{align}
    \begin{split}
        \hat{H}=\sum_i &-J\left(\hat{a}^\dagger_i\hat{a}_{i+1}+\hc\right)+2J\hat{a}^\dagger_i\hat{a}_{i}\\
        &+\frac{U}{2}\hat{a}^\dagger_i\hat{a}^\dagger_i\hat{a}_{i}\hat{a}_{i}\\
        &-J_\mathrm{I}\left(\hat{b}^\dagger_i\hat{b}_{i+1}+\hc\right)+2J_\mathrm{I}\hat{b}^\dagger_i\hat{b}_{i}\\
        &+U_\mathrm{I}\hat{b}^\dagger_i\hat{a}^\dagger_i\hat{a}_{i}\hat{b}_{i},
    \end{split}\label{eq:discretized_ham_hp}
\end{align}
where 
\begin{align}
    J=&\frac{1}{2m\Delta x^2},\qquad
    U=\frac{g}{\Delta x},\\
    J_\mathrm{I}=&\frac{1}{2M\Delta x^2},\qquad 
    U_\mathrm{I}=\frac{g_{IB}}{\Delta x}.
\end{align}
Here the terms containing $\hat{a}^\dagger_i\hat{a}_{i+1}$ ($\hat{b}^\dagger_i\hat{b}_{i+1}$) describe the hopping between lattice sites for the bosons (impurity) with hopping amplitude $J$ ($J_\mathrm{I}$).
The diagonal terms
$\sim \hat{a}^\dagger_i\hat{a}_{i}$ (or $\hat{b}^\dagger_i\hat{b}_{i}$) describe a local potential at site $i$ for the bosons (impurity). The strength of this potential is also given by the hopping amplitude $J$ ($J_\mathrm{I}$). It doesn't effect the dynamics of the system but needs to be taken into account when calculating the ground state energy.
The terms proportional to $\hat{a}^\dagger_i\hat{a}^\dagger_i\hat{a}_{i}\hat{a}_{i}$ ($\hat{b}^\dagger_i\hat{a}^\dagger_i\hat{a}_{i}\hat{b}_{i}$) relate to the interaction between the bosons (bosons and impurity). The strength of this interaction is given by $U$ ($U_\mathrm{I}$).
A graphical illustration of the terms of Hamiltonian \eqref{eq:discretized_ham_hp} is given in \figref{fig:sketch_discret_ham}a). 

%%%%%%%%%%%%%%%%%%%%%%%%%%%%%%%%%%%%%%%%%%%%%%
%%%%%%%%%%%%%%%%%%%%%%%%%%%%%%%%%%%%%%%%%%%%%

The discrete Hamiltonian is only a faithful approximation to the continuous model in the low-energy regime. Particles in a lattice possess a sinusoidal dispersion relation while free particles have a parabolic one (see \figref{fig:sketch_discret_ham}b). The two dispersion relations agree for up to a maximum energy of $\frac{1}{2m\Delta x^2}$ which relates to the hopping amplitude of the bosons $J$ in the discrete system. So the energy per lattice site caused by the interaction between the bosons needs to be lower then the hopping amplitude $J$.
This leads to the condition
\begin{align}
    \Tilde{n}_0U\ll J,
\end{align}
where $\Tilde{n}_0$ is the mean particle number per lattice site. In terms of the unitless Tonks parameter $\gamma$ this can be expressed as
\begin{align}
    \gamma\Tilde{n}_0^2\ll1.
\end{align}
Thus for larger Tonks parameter $\gamma$ the mean particle number per lattice site $\Tilde{n}_0$ needs to be kept low enough such that the discrete system remains a faithful approximation to the continues one.

In addition to the fact that the discrete system needs to be a faithful representation of the continues one, one has also to keep the compression of the DMRG in mind. We utilise the Julia language library ITensors \cite{itensor} for the implementation of the DMRG algorithm and set the bond dimension compression error to $10^{-6}$.  This is sufficiently small such that the state derived from the DMRG is a valid representation of the uncompressed state, but large enough that the system is still computable on a normal PC.

To gauge the effects of the impurity one can look at the polaron energy which gives the energy it takes to immerse an impurity in the Bose gas. 
\begin{align}
    E_P=E(\gib)-E(0)
\end{align}
where $E(0)$ is the ground-state energy of the system with $\gib=0$ and $E(\gib)$ is the corresponding ground-state energy for a finite boson-impurity interaction. The polaron energy increases with $\gib$ but saturates for large interaction strength once the impurity has displaced all of the condensate at its position.

%%%%%%%%%%%%%%%%%%%%%%%%%%%%%%%%%%%%%%%%%%%%%%
\begin{figure}[ht]
\centering
\includegraphics[width=0.8\columnwidth]{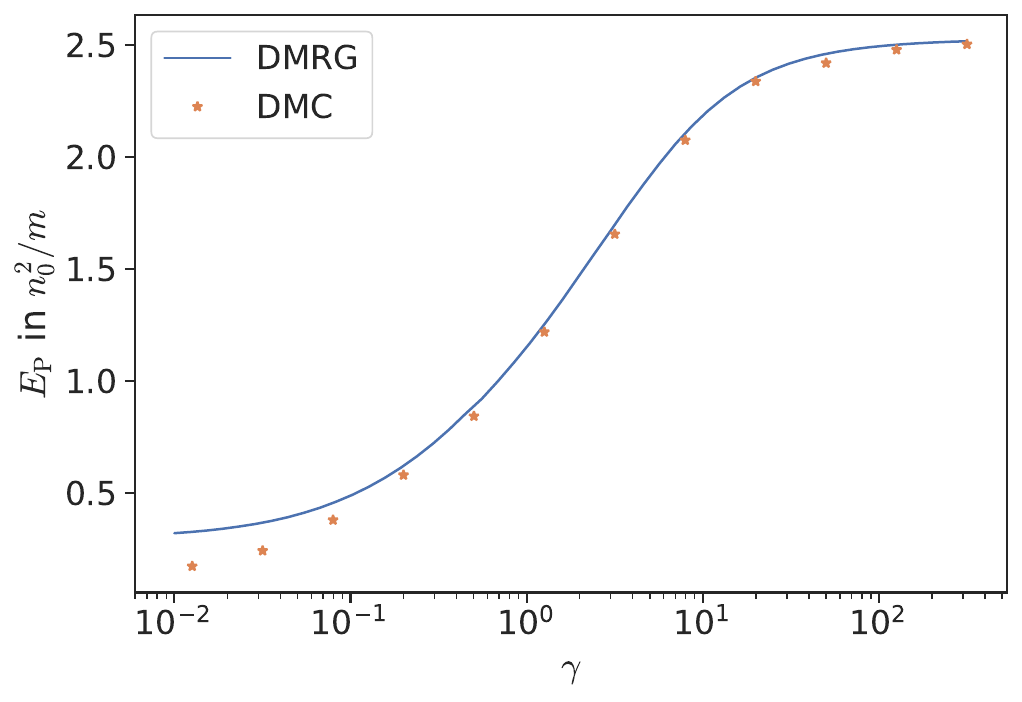}
\caption{Comparison of the saturated polaron energy ($\gib=\infty$) for an immobile impurity ($M=\infty$) in a Bose gas with $N=100$ particles from DMRG for OBC and diffusion quantum Monte-Carlo results (DMC) for PBC from \cite{Grusdt2017b}.
} \label{fig:DMRG_DMC} 
\end{figure}
%%%%%%%%%%%%%%%%%%%%%%%%%%%%%%%%%%%%%%%%%%%%%

In order to benchmark our DMRG code we calculated the saturated polaron energy ($\gib=\infty$) of an infinite-mass polaron 
$(M=\infty)$ localized in the center of a 1D quasi condensate with  $N=100$ bosons for varying strength of the boson-boson interaction, quantified by $\gamma$, and compared the results with those from diffusion quantum Monte Carlo (DMC) simulations taken from \cite{Grusdt2017b}, see \figref{fig:DMRG_DMC}. The DMC simulations were done for periodic boundary conditions, while the
DMRG data are for open boundary conditions, which explains the small difference for small value of $\gamma$.
Open boundary conditions cause half dark solitions to form on each system edge for $\gamma\ll1$ and Friedel oscillations  for $\gamma\gg1$ which increase the ground state energy of the system compared to periodic boundary conditions. This was mostly negated here by choosing a system size which is large compared to the characteristic length scale of the system, the healing length $\xi=1/\sqrt{2gn_0m}$ for $\gamma\ll1$ or the wavelength of Friedel oscillations $2k_F=2\pi n_0$ for $\gamma\gg1$.
Apart for very small $\gamma$ values one recognizes excellent agreement. For small $\gamma$ the polaron energies obtained with open boundary conditions (OBCs) as done in our DMR simulations are slightly larger than those obtained with periodic boundary conditions (PBCs) used in DMC as the spatial extend of the polaron, which is on the order of the healing length, becomes comparable to the system size:
\begin{align}
    \frac{L}{\xi}=\sqrt{2\gamma}N.
\end{align}

Let us finally comment on some techniques to improve the numerical simulations.
It turns out to be beneficial for the convergence of the DMRG algorithm to use a higher order representation of the second order derivatives. So 
\begin{align}
    \partial^2_x \hat{\phi}(x) &\approx \frac{-\frac{1}{12}\hat{a}_{i-2}+\frac{4}{3}\hat{a}_{i-1}-\frac{5}{2}\hat{a}_{i}+\frac{4}{3}\hat{a}_{i+1}-\frac{1}{12}\hat{a}_{i+2}}{\Delta x}.
\end{align}
By inserting these transformations in continuous Hamiltonian one arrives at
\begin{align}
    \begin{split}
        \hat{H}=\sum_i &\frac{1}{12}J\left(\hat{a}^\dagger_{i}\hat{a}_{i+2}+\hc\right)-\frac{4}{3}J\left(\hat{a}^\dagger_i\hat{a}_{i+1}+\hc\right)\\
        &+\frac{5}{2}J\left(\hat{a}^\dagger_i\hat{a}_{i}+\hc\right)+U\hat{a}^\dagger_i\hat{a}^\dagger_i\hat{a}_{i}\hat{a}_{i}\\
        &\frac{1}{12}J_I\left(\hat{b}^\dagger_{i}\hat{b}_{i+2}+\hc\right)-\frac{4}{3}J_I\left(\hat{b}^\dagger_i\hat{b}_{i+1}+\hc\right)\\
        &+\frac{5}{2}J_I\left(\hat{b}^\dagger_i\hat{b}_{i}+\hc\right)+U_I\hat{b}^\dagger_i\hat{a}^\dagger_i\hat{a}_{i}\hat{b}_{i}.
    \end{split}\label{eq:diskret_ham_hp}
\end{align}
As compared to a lattice Hamiltonian derived from \eqref{eq:discret_second_deriv}, the above expression \eqref{eq:diskret_ham_hp} contains longer-range hopping terms. These help the DMRG algorithm to establish correlations and therefore the algorithm is able to reach the ground state faster.

%%%%%%%%%%%%%%%%%%%%%%%%%%%%%%%%%%%%%%%%%%%%%%%
\subsection{Comparison of polaron-polaron interaction potential to mean-field result}
%%%%%%%%%%%%%%%%%%%%%%%%%%%%%%%%%%%%%%%%%%%%%%%

In the case of a weakly interacting Bose gas with $\gamma \ll 1$ the interaction potential between two Bose polarons in Born-Oppenheimer approximation can be determined semi-analytical \cite{Will2021}.  One finds for repulsive interactions
\begin{align}
	& V(r)
	= g n_0^2 r \left(\frac{1}{2}- \frac{4+2\nu}{3(\nu+1)^2} \right)+	
	\frac{4}{3} \frac{g n_0^2 \bar{\xi}}{\sqrt{1+\nu}} \Bigg\{
\sqrt{2\nu+2}
	 \notag\\
	 &+ 2 \, \text{E} \big(\am(u,\nu),\nu\big) - \frac{\sqrt{{\nu}}^3}{1+v} \cd(u,\nu)^{3} \left[1+\sqrt{{\nu}} \; \sn(u,\nu) \right]
	\notag \\
	& - \sqrt{{\nu}} \;	\cd(u,\nu) \left[ \frac{3}{2} + \frac{1+2 \nu}{1+\nu}\sqrt{\tilde{\nu}} \; \sn(u,\nu)
	\right]
	\Bigg\}, \label{eq:V_r}
\end{align}
where % $n_0 =\vert\phi_0\vert^2$ is the mean-field density far away from the impurities,
$u ={r}/(2\bar{\xi} \sqrt{1+\nu})$ is a normalized distance, $\text{E}(x,\nu)$ is the incomplete elliptic integral of the second kind, $\cd(x,\nu)$ and $\sn(x,\nu)$ are Jacobi elliptic functions and $\am(x,\nu)$ is the amplitude of these functions \cite{Lawden1989}. The dimensionless parameter $\nu = \nu(r,\eta)$ with $|\nu|<1$ is given implicitly by 
\begin{align}
2 \frac{|\eta|}{n_0 \xit} \frac{\sqrt{{\nu}(\nu+1)}}{(1-\nu)} \; &\cn \left(u, \nu \right)\, \dn \left(u, \nu \right)= \left[ 1 + \sqrt{{\nu}} \,\sn \left(u, \nu \right) \right]^2, 
\nonumber
\end{align}
involving the Jacobi elliptic sn, cn, and dn functions and $\eta = \gib / g > 0$.  In general, this equation has several solutions, however the physically relevant one is that with the largest $\nu$.

\begin{figure}[ht]
\centering
\includegraphics[width=0.9\columnwidth]{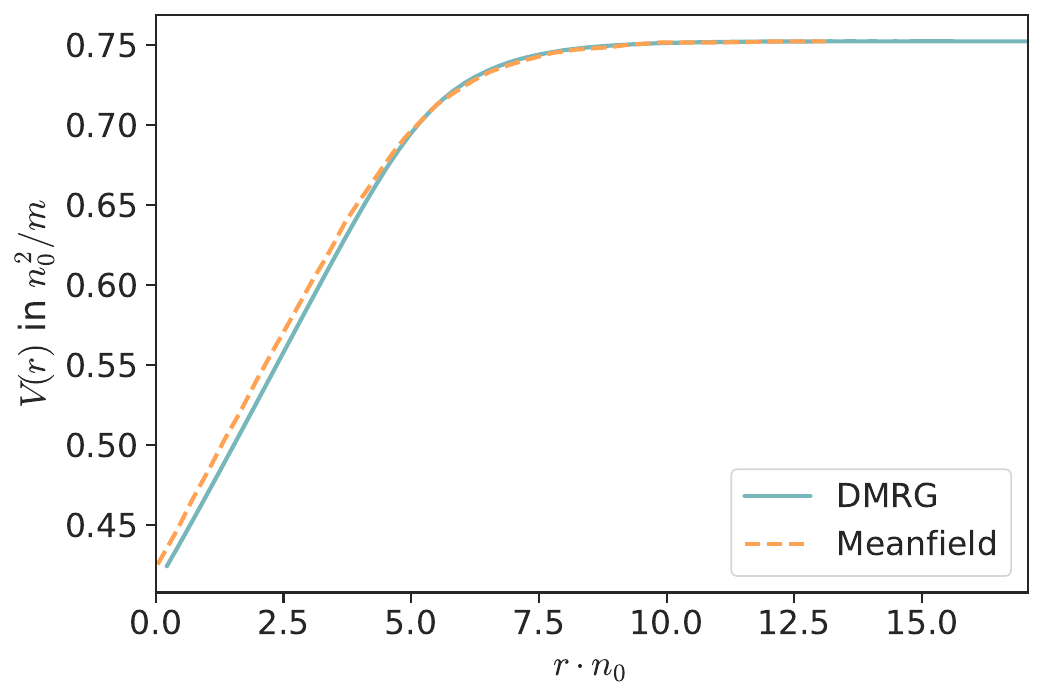}
\caption{Comparison of polaron-polaron interaction potential from Meanfield calculations \cite{Will2021} (dashed lines) and from DMRG simulations (solid lines) for a boson-impurity interaction strength $\gib\frac{m}{n_0}=1.25$ and a weakly interacting Bose gas with $\gamma=1/8$ and $N=80$ particles. 
} \label{fig:bipolaron-weak-comparison-Martin} 
\end{figure}

In \figref{fig:bipolaron-weak-comparison-Martin} we have shown a comparison of the DMRG results (full line) with the above prediction (dashed line).

%%%%%%%%%%%%%%%%%%%%%%%%%%%%%%%%%%%%%%%%%%%%%%
%%%%%%%%%%%%%%%%%%%%%%%%%%%%%%%%%%%%%%%%%%%%% 

%
\bibliography{library.bib}

%apsrev4-2.bst 2019-01-14 (MD) hand-edited version of apsrev4-1.bst
%Control: key (0)
%Control: author (8) initials jnrlst
%Control: editor formatted (1) identically to author
%Control: production of article title (0) allowed
%Control: page (0) single
%Control: year (1) truncated
%Control: production of eprint (0) enabled
\begin{thebibliography}{56}%
\makeatletter
\providecommand \@ifxundefined [1]{%
 \@ifx{#1\undefined}
}%
\providecommand \@ifnum [1]{%
 \ifnum #1\expandafter \@firstoftwo
 \else \expandafter \@secondoftwo
 \fi
}%
\providecommand \@ifx [1]{%
 \ifx #1\expandafter \@firstoftwo
 \else \expandafter \@secondoftwo
 \fi
}%
\providecommand \natexlab [1]{#1}%
\providecommand \enquote  [1]{``#1''}%
\providecommand \bibnamefont  [1]{#1}%
\providecommand \bibfnamefont [1]{#1}%
\providecommand \citenamefont [1]{#1}%
\providecommand \href@noop [0]{\@secondoftwo}%
\providecommand \href [0]{\begingroup \@sanitize@url \@href}%
\providecommand \@href[1]{\@@startlink{#1}\@@href}%
\providecommand \@@href[1]{\endgroup#1\@@endlink}%
\providecommand \@sanitize@url [0]{\catcode `\\12\catcode `\$12\catcode `\&12\catcode `\#12\catcode `\^12\catcode `\_12\catcode `\%12\relax}%
\providecommand \@@startlink[1]{}%
\providecommand \@@endlink[0]{}%
\providecommand \url  [0]{\begingroup\@sanitize@url \@url }%
\providecommand \@url [1]{\endgroup\@href {#1}{\urlprefix }}%
\providecommand \urlprefix  [0]{URL }%
\providecommand \Eprint [0]{\href }%
\providecommand \doibase [0]{https://doi.org/}%
\providecommand \selectlanguage [0]{\@gobble}%
\providecommand \bibinfo  [0]{\@secondoftwo}%
\providecommand \bibfield  [0]{\@secondoftwo}%
\providecommand \translation [1]{[#1]}%
\providecommand \BibitemOpen [0]{}%
\providecommand \bibitemStop [0]{}%
\providecommand \bibitemNoStop [0]{.\EOS\space}%
\providecommand \EOS [0]{\spacefactor3000\relax}%
\providecommand \BibitemShut  [1]{\csname bibitem#1\endcsname}%
\let\auto@bib@innerbib\@empty
%</preamble>
\bibitem [{\citenamefont {Alexandrov}\ and\ \citenamefont {Devreese}(2010)}]{Alexandrov2010}%
  \BibitemOpen
  \bibfield  {author} {\bibinfo {author} {\bibfnamefont {A.~S.}\ \bibnamefont {Alexandrov}}\ and\ \bibinfo {author} {\bibfnamefont {J.~T.}\ \bibnamefont {Devreese}},\ }\href {https://doi.org/10.1007/978-3-642-01896-1} {\emph {\bibinfo {title} {{Advances in Polaron Physics}}}},\ Vol.\ \bibinfo {volume} {159}\ (\bibinfo  {publisher} {Springer-Verlag, Berlin},\ \bibinfo {year} {2010})\BibitemShut {NoStop}%
\bibitem [{\citenamefont {Ruderman}\ and\ \citenamefont {Kittel}(1954)}]{Ruderman1954}%
  \BibitemOpen
  \bibfield  {author} {\bibinfo {author} {\bibfnamefont {M.~A.}\ \bibnamefont {Ruderman}}\ and\ \bibinfo {author} {\bibfnamefont {C.}~\bibnamefont {Kittel}},\ }\bibfield  {title} {\bibinfo {title} {Indirect exchange coupling of nuclear magnetic moments by conduction electrons},\ }\href {https://doi.org/10.1103/PhysRev.96.99} {\bibfield  {journal} {\bibinfo  {journal} {Phys. Rev.}\ }\textbf {\bibinfo {volume} {96}},\ \bibinfo {pages} {99} (\bibinfo {year} {1954})}\BibitemShut {NoStop}%
\bibitem [{\citenamefont {Kasuya}(1956)}]{Kasuya1956}%
  \BibitemOpen
  \bibfield  {author} {\bibinfo {author} {\bibfnamefont {T.}~\bibnamefont {Kasuya}},\ }\bibfield  {title} {\bibinfo {title} {{A Theory of Metallic Ferro- and Antiferromagnetism on Zener's Model}},\ }\href {https://doi.org/10.1143/PTP.16.45} {\bibfield  {journal} {\bibinfo  {journal} {Progress of Theoretical Physics}\ }\textbf {\bibinfo {volume} {16}},\ \bibinfo {pages} {45} (\bibinfo {year} {1956})}\BibitemShut {NoStop}%
\bibitem [{\citenamefont {Yosida}(1957)}]{Yosida1957}%
  \BibitemOpen
  \bibfield  {author} {\bibinfo {author} {\bibfnamefont {K.}~\bibnamefont {Yosida}},\ }\bibfield  {title} {\bibinfo {title} {Magnetic properties of cu-mn alloys},\ }\href {https://doi.org/10.1103/PhysRev.106.893} {\bibfield  {journal} {\bibinfo  {journal} {Phys. Rev.}\ }\textbf {\bibinfo {volume} {106}},\ \bibinfo {pages} {893} (\bibinfo {year} {1957})}\BibitemShut {NoStop}%
\bibitem [{\citenamefont {Cooper}(1956)}]{Cooper1956}%
  \BibitemOpen
  \bibfield  {author} {\bibinfo {author} {\bibfnamefont {L.~N.}\ \bibnamefont {Cooper}},\ }\bibfield  {title} {\bibinfo {title} {Bound electron pairs in a degenerate fermi gas},\ }\href {https://doi.org/10.1103/PhysRev.104.1189} {\bibfield  {journal} {\bibinfo  {journal} {Phys. Rev.}\ }\textbf {\bibinfo {volume} {104}},\ \bibinfo {pages} {1189} (\bibinfo {year} {1956})}\BibitemShut {NoStop}%
\bibitem [{\citenamefont {Landau}(1933)}]{Landau1933}%
  \BibitemOpen
  \bibfield  {author} {\bibinfo {author} {\bibfnamefont {L.}~\bibnamefont {Landau}},\ }\bibfield  {title} {\bibinfo {title} {{{\"{U}}ber die Bewegung der Elektronen in Kristalgitter}},\ }\href@noop {} {\bibfield  {journal} {\bibinfo  {journal} {Phys. Z. Sowjetunion}\ }\textbf {\bibinfo {volume} {3}},\ \bibinfo {pages} {644} (\bibinfo {year} {1933})}\BibitemShut {NoStop}%
\bibitem [{\citenamefont {Pekar}(1946)}]{Pekar1946}%
  \BibitemOpen
  \bibfield  {author} {\bibinfo {author} {\bibfnamefont {S.~I.}\ \bibnamefont {Pekar}},\ }\bibfield  {title} {\bibinfo {title} {Effective mass of a polaron},\ }\href@noop {} {\bibfield  {journal} {\bibinfo  {journal} {Zh. Eksp. Teor. Fiz.}\ }\textbf {\bibinfo {volume} {16}},\ \bibinfo {pages} {335} (\bibinfo {year} {1946})}\BibitemShut {NoStop}%
\bibitem [{\citenamefont {Fr{\"{o}}hlich}(1954)}]{Frohlich1954}%
  \BibitemOpen
  \bibfield  {author} {\bibinfo {author} {\bibfnamefont {H.}~\bibnamefont {Fr{\"{o}}hlich}},\ }\bibfield  {title} {\bibinfo {title} {{Electrons in lattice fields}},\ }\href {https://doi.org/10.1080/00018735400101213} {\bibfield  {journal} {\bibinfo  {journal} {Adv. Phys.}\ }\textbf {\bibinfo {volume} {3}},\ \bibinfo {pages} {325} (\bibinfo {year} {1954})}\BibitemShut {NoStop}%
\bibitem [{\citenamefont {Grusdt}\ \emph {et~al.}(2017)\citenamefont {Grusdt}, \citenamefont {Astrakharchik},\ and\ \citenamefont {Demler}}]{Grusdt2017b}%
  \BibitemOpen
  \bibfield  {author} {\bibinfo {author} {\bibfnamefont {F.}~\bibnamefont {Grusdt}}, \bibinfo {author} {\bibfnamefont {G.~E.}\ \bibnamefont {Astrakharchik}},\ and\ \bibinfo {author} {\bibfnamefont {E.}~\bibnamefont {Demler}},\ }\bibfield  {title} {\bibinfo {title} {Bose polarons in ultracold atoms in one dimension: beyond the fröhlich paradigm},\ }\href {https://doi.org/10.1088/1367-2630/aa8a2e} {\bibfield  {journal} {\bibinfo  {journal} {New Journal of Physics}\ }\textbf {\bibinfo {volume} {19}},\ \bibinfo {pages} {103035} (\bibinfo {year} {2017})}\BibitemShut {NoStop}%
\bibitem [{\citenamefont {Lee}\ \emph {et~al.}(1953)\citenamefont {Lee}, \citenamefont {Low},\ and\ \citenamefont {Pines}}]{Lee1953}%
  \BibitemOpen
  \bibfield  {author} {\bibinfo {author} {\bibfnamefont {T.~D.}\ \bibnamefont {Lee}}, \bibinfo {author} {\bibfnamefont {F.~E.}\ \bibnamefont {Low}},\ and\ \bibinfo {author} {\bibfnamefont {D.}~\bibnamefont {Pines}},\ }\bibfield  {title} {\bibinfo {title} {The motion of slow electrons in a polar crystal},\ }\href {https://doi.org/10.1103/PhysRev.90.297} {\bibfield  {journal} {\bibinfo  {journal} {Phys. Rev.}\ }\textbf {\bibinfo {volume} {90}},\ \bibinfo {pages} {297} (\bibinfo {year} {1953})}\BibitemShut {NoStop}%
\bibitem [{\citenamefont {Volosniev}\ and\ \citenamefont {Hammer}(2017)}]{Volosniev2017}%
  \BibitemOpen
  \bibfield  {author} {\bibinfo {author} {\bibfnamefont {A.~G.}\ \bibnamefont {Volosniev}}\ and\ \bibinfo {author} {\bibfnamefont {H.-W.}\ \bibnamefont {Hammer}},\ }\bibfield  {title} {\bibinfo {title} {Analytical approach to the bose-polaron problem in one dimension},\ }\href {https://doi.org/10.1103/PhysRevA.96.031601} {\bibfield  {journal} {\bibinfo  {journal} {Phys. Rev. A}\ }\textbf {\bibinfo {volume} {96}},\ \bibinfo {pages} {031601} (\bibinfo {year} {2017})}\BibitemShut {NoStop}%
\bibitem [{\citenamefont {Jager}\ \emph {et~al.}(2020)\citenamefont {Jager}, \citenamefont {Barnett}, \citenamefont {Will},\ and\ \citenamefont {Fleischhauer}}]{Jager2020}%
  \BibitemOpen
  \bibfield  {author} {\bibinfo {author} {\bibfnamefont {J.}~\bibnamefont {Jager}}, \bibinfo {author} {\bibfnamefont {R.}~\bibnamefont {Barnett}}, \bibinfo {author} {\bibfnamefont {M.}~\bibnamefont {Will}},\ and\ \bibinfo {author} {\bibfnamefont {M.}~\bibnamefont {Fleischhauer}},\ }\bibfield  {title} {\bibinfo {title} {Strong-coupling bose polarons in one dimension: Condensate deformation and modified bogoliubov phonons},\ }\href {https://doi.org/10.1103/PhysRevResearch.2.033142} {\bibfield  {journal} {\bibinfo  {journal} {Phys. Rev. Res.}\ }\textbf {\bibinfo {volume} {2}},\ \bibinfo {pages} {033142} (\bibinfo {year} {2020})}\BibitemShut {NoStop}%
\bibitem [{\citenamefont {Will}\ \emph {et~al.}(2021)\citenamefont {Will}, \citenamefont {Astrakharchik},\ and\ \citenamefont {Fleischhauer}}]{Will2021}%
  \BibitemOpen
  \bibfield  {author} {\bibinfo {author} {\bibfnamefont {M.}~\bibnamefont {Will}}, \bibinfo {author} {\bibfnamefont {G.~E.}\ \bibnamefont {Astrakharchik}},\ and\ \bibinfo {author} {\bibfnamefont {M.}~\bibnamefont {Fleischhauer}},\ }\bibfield  {title} {\bibinfo {title} {Polaron interactions and bipolarons in one-dimensional bose gases in the strong coupling regime},\ }\href {https://doi.org/10.1103/PhysRevLett.127.103401} {\bibfield  {journal} {\bibinfo  {journal} {Phys. Rev. Lett.}\ }\textbf {\bibinfo {volume} {127}},\ \bibinfo {pages} {103401} (\bibinfo {year} {2021})}\BibitemShut {NoStop}%
\bibitem [{\citenamefont {Will}\ and\ \citenamefont {Fleischhauer}(2023)}]{will2023dynamics}%
  \BibitemOpen
  \bibfield  {author} {\bibinfo {author} {\bibfnamefont {M.}~\bibnamefont {Will}}\ and\ \bibinfo {author} {\bibfnamefont {M.}~\bibnamefont {Fleischhauer}},\ }\bibfield  {title} {\bibinfo {title} {Dynamics of polaron formation in 1d bose gases in the strong-coupling regime},\ }\href@noop {} {\bibfield  {journal} {\bibinfo  {journal} {New Journal of Physics}\ }\textbf {\bibinfo {volume} {25}},\ \bibinfo {pages} {083043} (\bibinfo {year} {2023})}\BibitemShut {NoStop}%
\bibitem [{\citenamefont {Panochko}\ and\ \citenamefont {Pastukhov}(2019)}]{Panochko2019}%
  \BibitemOpen
  \bibfield  {author} {\bibinfo {author} {\bibfnamefont {G.}~\bibnamefont {Panochko}}\ and\ \bibinfo {author} {\bibfnamefont {V.}~\bibnamefont {Pastukhov}},\ }\bibfield  {title} {\bibinfo {title} {Mean-field construction for spectrum of one-dimensional bose polaron},\ }\href {https://doi.org/https://doi.org/10.1016/j.aop.2019.167933} {\bibfield  {journal} {\bibinfo  {journal} {Annals of Physics}\ }\textbf {\bibinfo {volume} {409}},\ \bibinfo {pages} {167933} (\bibinfo {year} {2019})}\BibitemShut {NoStop}%
\bibitem [{\citenamefont {Timmermans}(1998)}]{timmermans1998phase}%
  \BibitemOpen
  \bibfield  {author} {\bibinfo {author} {\bibfnamefont {E.}~\bibnamefont {Timmermans}},\ }\bibfield  {title} {\bibinfo {title} {Phase separation of bose-einstein condensates},\ }\href@noop {} {\bibfield  {journal} {\bibinfo  {journal} {Physical review letters}\ }\textbf {\bibinfo {volume} {81}},\ \bibinfo {pages} {5718} (\bibinfo {year} {1998})}\BibitemShut {NoStop}%
\bibitem [{\citenamefont {Lee}\ and\ \citenamefont {Gunn}(1992)}]{lee1992polarons}%
  \BibitemOpen
  \bibfield  {author} {\bibinfo {author} {\bibfnamefont {D.}~\bibnamefont {Lee}}\ and\ \bibinfo {author} {\bibfnamefont {J.}~\bibnamefont {Gunn}},\ }\bibfield  {title} {\bibinfo {title} {Polarons and bose decondensation: A self-trapping approach},\ }\href@noop {} {\bibfield  {journal} {\bibinfo  {journal} {Physical Review B}\ }\textbf {\bibinfo {volume} {46}},\ \bibinfo {pages} {301} (\bibinfo {year} {1992})}\BibitemShut {NoStop}%
\bibitem [{\citenamefont {Cucchietti}\ and\ \citenamefont {Timmermans}(2006{\natexlab{a}})}]{cucchietti2006strong}%
  \BibitemOpen
  \bibfield  {author} {\bibinfo {author} {\bibfnamefont {F.}~\bibnamefont {Cucchietti}}\ and\ \bibinfo {author} {\bibfnamefont {E.}~\bibnamefont {Timmermans}},\ }\bibfield  {title} {\bibinfo {title} {Strong-coupling polarons in dilute gas bose-einstein condensates},\ }\href@noop {} {\bibfield  {journal} {\bibinfo  {journal} {Physical review letters}\ }\textbf {\bibinfo {volume} {96}},\ \bibinfo {pages} {210401} (\bibinfo {year} {2006}{\natexlab{a}})}\BibitemShut {NoStop}%
\bibitem [{\citenamefont {Sacha}\ and\ \citenamefont {Timmermans}(2006)}]{sacha2006self}%
  \BibitemOpen
  \bibfield  {author} {\bibinfo {author} {\bibfnamefont {K.}~\bibnamefont {Sacha}}\ and\ \bibinfo {author} {\bibfnamefont {E.}~\bibnamefont {Timmermans}},\ }\bibfield  {title} {\bibinfo {title} {Self-localized impurities embedded in a one-dimensional bose-einstein condensate and their quantum fluctuations},\ }\href@noop {} {\bibfield  {journal} {\bibinfo  {journal} {Physical Review A—Atomic, Molecular, and Optical Physics}\ }\textbf {\bibinfo {volume} {73}},\ \bibinfo {pages} {063604} (\bibinfo {year} {2006})}\BibitemShut {NoStop}%
\bibitem [{\citenamefont {Kalas}\ and\ \citenamefont {Blume}(2006)}]{kalas2006interaction}%
  \BibitemOpen
  \bibfield  {author} {\bibinfo {author} {\bibfnamefont {R.~M.}\ \bibnamefont {Kalas}}\ and\ \bibinfo {author} {\bibfnamefont {D.}~\bibnamefont {Blume}},\ }\bibfield  {title} {\bibinfo {title} {Interaction-induced localization of an impurity in a trapped bose-einstein condensate},\ }\href@noop {} {\bibfield  {journal} {\bibinfo  {journal} {Physical Review A—Atomic, Molecular, and Optical Physics}\ }\textbf {\bibinfo {volume} {73}},\ \bibinfo {pages} {043608} (\bibinfo {year} {2006})}\BibitemShut {NoStop}%
\bibitem [{\citenamefont {Bruderer}\ \emph {et~al.}(2008{\natexlab{a}})\citenamefont {Bruderer}, \citenamefont {Bao},\ and\ \citenamefont {Jaksch}}]{bruderer2008self}%
  \BibitemOpen
  \bibfield  {author} {\bibinfo {author} {\bibfnamefont {M.}~\bibnamefont {Bruderer}}, \bibinfo {author} {\bibfnamefont {W.}~\bibnamefont {Bao}},\ and\ \bibinfo {author} {\bibfnamefont {D.}~\bibnamefont {Jaksch}},\ }\bibfield  {title} {\bibinfo {title} {Self-trapping of impurities in bose-einstein condensates: Strong attractive and repulsive coupling},\ }\href@noop {} {\bibfield  {journal} {\bibinfo  {journal} {Europhysics Letters}\ }\textbf {\bibinfo {volume} {82}},\ \bibinfo {pages} {30004} (\bibinfo {year} {2008}{\natexlab{a}})}\BibitemShut {NoStop}%
\bibitem [{\citenamefont {Blinova}\ \emph {et~al.}(2013)\citenamefont {Blinova}, \citenamefont {Boshier},\ and\ \citenamefont {Timmermans}}]{blinova2013two}%
  \BibitemOpen
  \bibfield  {author} {\bibinfo {author} {\bibfnamefont {A.}~\bibnamefont {Blinova}}, \bibinfo {author} {\bibfnamefont {M.}~\bibnamefont {Boshier}},\ and\ \bibinfo {author} {\bibfnamefont {E.}~\bibnamefont {Timmermans}},\ }\bibfield  {title} {\bibinfo {title} {Two polaron flavors of the bose-einstein condensate impurity},\ }\href@noop {} {\bibfield  {journal} {\bibinfo  {journal} {Physical Review A—Atomic, Molecular, and Optical Physics}\ }\textbf {\bibinfo {volume} {88}},\ \bibinfo {pages} {053610} (\bibinfo {year} {2013})}\BibitemShut {NoStop}%
\bibitem [{\citenamefont {Brewer}\ \emph {et~al.}(1966)\citenamefont {Brewer}, \citenamefont {of~Sussex}, \citenamefont {of~Pure},\ and\ \citenamefont {Physics}}]{brewer1966quantum}%
  \BibitemOpen
  \bibfield  {author} {\bibinfo {author} {\bibfnamefont {D.}~\bibnamefont {Brewer}}, \bibinfo {author} {\bibfnamefont {U.}~\bibnamefont {of~Sussex}}, \bibinfo {author} {\bibfnamefont {I.~U.}\ \bibnamefont {of~Pure}},\ and\ \bibinfo {author} {\bibfnamefont {A.}~\bibnamefont {Physics}},\ }\href {https://books.google.de/books?id=yc4NAQAAIAAJ} {\emph {\bibinfo {title} {Quantum Fluids: Proceedings}}}\ (\bibinfo  {publisher} {North-Holland Publishing Company},\ \bibinfo {year} {1966})\BibitemShut {NoStop}%
\bibitem [{\citenamefont {Hernandez}(1991)}]{hernandez1991electron}%
  \BibitemOpen
  \bibfield  {author} {\bibinfo {author} {\bibfnamefont {J.~P.}\ \bibnamefont {Hernandez}},\ }\bibfield  {title} {\bibinfo {title} {Electron self-trapping in liquids and dense gases},\ }\href@noop {} {\bibfield  {journal} {\bibinfo  {journal} {Reviews of modern physics}\ }\textbf {\bibinfo {volume} {63}},\ \bibinfo {pages} {675} (\bibinfo {year} {1991})}\BibitemShut {NoStop}%
\bibitem [{\citenamefont {Zschetzsche}\ and\ \citenamefont {Zillich}(2024)}]{zschetzsche2024suppression}%
  \BibitemOpen
  \bibfield  {author} {\bibinfo {author} {\bibfnamefont {L.}~\bibnamefont {Zschetzsche}}\ and\ \bibinfo {author} {\bibfnamefont {R.~E.}\ \bibnamefont {Zillich}},\ }\bibfield  {title} {\bibinfo {title} {Suppression of polaron self-localization by correlations},\ }\href@noop {} {\bibfield  {journal} {\bibinfo  {journal} {Physical Review Research}\ }\textbf {\bibinfo {volume} {6}},\ \bibinfo {pages} {023137} (\bibinfo {year} {2024})}\BibitemShut {NoStop}%
\bibitem [{\citenamefont {Schollwöck}(2011)}]{schollwock}%
  \BibitemOpen
  \bibfield  {author} {\bibinfo {author} {\bibfnamefont {U.}~\bibnamefont {Schollwöck}},\ }\bibfield  {title} {\bibinfo {title} {The density-matrix renormalization group in the age of matrix product states},\ }\href {https://doi.org/https://doi.org/10.1016/j.aop.2010.09.012} {\bibfield  {journal} {\bibinfo  {journal} {Annals of Physics}\ }\textbf {\bibinfo {volume} {326}},\ \bibinfo {pages} {96} (\bibinfo {year} {2011})},\ \bibinfo {note} {january 2011 Special Issue}\BibitemShut {NoStop}%
\bibitem [{\citenamefont {Fishman}\ \emph {et~al.}(2022)\citenamefont {Fishman}, \citenamefont {White},\ and\ \citenamefont {Stoudenmire}}]{itensor}%
  \BibitemOpen
  \bibfield  {author} {\bibinfo {author} {\bibfnamefont {M.}~\bibnamefont {Fishman}}, \bibinfo {author} {\bibfnamefont {S.~R.}\ \bibnamefont {White}},\ and\ \bibinfo {author} {\bibfnamefont {E.~M.}\ \bibnamefont {Stoudenmire}},\ }\bibfield  {title} {\bibinfo {title} {{The ITensor Software Library for Tensor Network Calculations}},\ }\href {https://doi.org/10.21468/SciPostPhysCodeb.4} {\bibfield  {journal} {\bibinfo  {journal} {SciPost Phys. Codebases}\ ,\ \bibinfo {pages} {4}} (\bibinfo {year} {2022})}\BibitemShut {NoStop}%
\bibitem [{\citenamefont {Recati}\ \emph {et~al.}(2005)\citenamefont {Recati}, \citenamefont {Fuchs}, \citenamefont {Peca},\ and\ \citenamefont {Zwerger}}]{recati2005casimir}%
  \BibitemOpen
  \bibfield  {author} {\bibinfo {author} {\bibfnamefont {A.}~\bibnamefont {Recati}}, \bibinfo {author} {\bibfnamefont {J.-N.}\ \bibnamefont {Fuchs}}, \bibinfo {author} {\bibfnamefont {C.~S.}\ \bibnamefont {Peca}},\ and\ \bibinfo {author} {\bibfnamefont {W.}~\bibnamefont {Zwerger}},\ }\bibfield  {title} {\bibinfo {title} {Casimir forces between defects in one-dimensional quantum liquids},\ }\href@noop {} {\bibfield  {journal} {\bibinfo  {journal} {Physical Review A—Atomic, Molecular, and Optical Physics}\ }\textbf {\bibinfo {volume} {72}},\ \bibinfo {pages} {023616} (\bibinfo {year} {2005})}\BibitemShut {NoStop}%
\bibitem [{\citenamefont {Fuchs}\ \emph {et~al.}(2007)\citenamefont {Fuchs}, \citenamefont {Recati},\ and\ \citenamefont {Zwerger}}]{fuchs2007oscillating}%
  \BibitemOpen
  \bibfield  {author} {\bibinfo {author} {\bibfnamefont {J.}~\bibnamefont {Fuchs}}, \bibinfo {author} {\bibfnamefont {A.}~\bibnamefont {Recati}},\ and\ \bibinfo {author} {\bibfnamefont {W.}~\bibnamefont {Zwerger}},\ }\bibfield  {title} {\bibinfo {title} {Oscillating casimir force between impurities in one-dimensional fermi liquids},\ }\href@noop {} {\bibfield  {journal} {\bibinfo  {journal} {Physical Review A—Atomic, Molecular, and Optical Physics}\ }\textbf {\bibinfo {volume} {75}},\ \bibinfo {pages} {043615} (\bibinfo {year} {2007})}\BibitemShut {NoStop}%
\bibitem [{\citenamefont {Mermin}\ and\ \citenamefont {Wagner}(1966)}]{1D_2D_bec1}%
  \BibitemOpen
  \bibfield  {author} {\bibinfo {author} {\bibfnamefont {N.~D.}\ \bibnamefont {Mermin}}\ and\ \bibinfo {author} {\bibfnamefont {H.}~\bibnamefont {Wagner}},\ }\bibfield  {title} {\bibinfo {title} {Absence of ferromagnetism or antiferromagnetism in one- or two-dimensional isotropic heisenberg models},\ }\href {https://doi.org/10.1103/PhysRevLett.17.1133} {\bibfield  {journal} {\bibinfo  {journal} {Phys. Rev. Lett.}\ }\textbf {\bibinfo {volume} {17}},\ \bibinfo {pages} {1133} (\bibinfo {year} {1966})}\BibitemShut {NoStop}%
\bibitem [{\citenamefont {Hohenberg}(1967)}]{1D_2D_bec2}%
  \BibitemOpen
  \bibfield  {author} {\bibinfo {author} {\bibfnamefont {P.~C.}\ \bibnamefont {Hohenberg}},\ }\bibfield  {title} {\bibinfo {title} {Existence of long-range order in one and two dimensions},\ }\href {https://doi.org/10.1103/PhysRev.158.383} {\bibfield  {journal} {\bibinfo  {journal} {Phys. Rev.}\ }\textbf {\bibinfo {volume} {158}},\ \bibinfo {pages} {383} (\bibinfo {year} {1967})}\BibitemShut {NoStop}%
\bibitem [{\citenamefont {Girardeau}(1960)}]{Girardeau1960}%
  \BibitemOpen
  \bibfield  {author} {\bibinfo {author} {\bibfnamefont {M.}~\bibnamefont {Girardeau}},\ }\bibfield  {title} {\bibinfo {title} {{Relationship between Systems of Impenetrable Bosons and Fermions in One Dimension}},\ }\href {https://doi.org/10.1063/1.1703687} {\bibfield  {journal} {\bibinfo  {journal} {Journal of Mathematical Physics}\ }\textbf {\bibinfo {volume} {1}},\ \bibinfo {pages} {516} (\bibinfo {year} {1960})},\ \Eprint {https://arxiv.org/abs/https://pubs.aip.org/aip/jmp/article-pdf/1/6/516/19055341/516\_1\_online.pdf} {https://pubs.aip.org/aip/jmp/article-pdf/1/6/516/19055341/516\_1\_online.pdf} \BibitemShut {NoStop}%
\bibitem [{\citenamefont {Cucchietti}\ and\ \citenamefont {Timmermans}(2006{\natexlab{b}})}]{Cucchietti2006}%
  \BibitemOpen
  \bibfield  {author} {\bibinfo {author} {\bibfnamefont {F.~M.}\ \bibnamefont {Cucchietti}}\ and\ \bibinfo {author} {\bibfnamefont {E.}~\bibnamefont {Timmermans}},\ }\bibfield  {title} {\bibinfo {title} {{Strong-coupling polarons in dilute gas Bose-Einstein condensates}},\ }\bibfield  {journal} {\bibinfo  {journal} {Phys. Rev. Lett.}\ }\textbf {\bibinfo {volume} {96}},\ \href {https://doi.org/10.1103/PhysRevLett.96.210401} {10.1103/PhysRevLett.96.210401} (\bibinfo {year} {2006}{\natexlab{b}})\BibitemShut {NoStop}%
\bibitem [{\citenamefont {Bruderer}\ \emph {et~al.}(2008{\natexlab{b}})\citenamefont {Bruderer}, \citenamefont {Bao},\ and\ \citenamefont {Jaksch}}]{Bruderer-EPL2008}%
  \BibitemOpen
  \bibfield  {author} {\bibinfo {author} {\bibfnamefont {M.}~\bibnamefont {Bruderer}}, \bibinfo {author} {\bibfnamefont {W.}~\bibnamefont {Bao}},\ and\ \bibinfo {author} {\bibfnamefont {D.}~\bibnamefont {Jaksch}},\ }\bibfield  {title} {\bibinfo {title} {Self-trapping of impurities in bose-einstein condensates: Strong attractive and repulsive coupling},\ }\href {https://doi.org/10.1209/0295-5075/82/30004} {\bibfield  {journal} {\bibinfo  {journal} {Europhysics Letters}\ }\textbf {\bibinfo {volume} {82}},\ \bibinfo {pages} {30004} (\bibinfo {year} {2008}{\natexlab{b}})}\BibitemShut {NoStop}%
\bibitem [{\citenamefont {Nieto}(1978)}]{nieto1978exact}%
  \BibitemOpen
  \bibfield  {author} {\bibinfo {author} {\bibfnamefont {M.~M.}\ \bibnamefont {Nieto}},\ }\bibfield  {title} {\bibinfo {title} {Exact wave-function normalization constants for the b 0 tanh z- u 0 cosh- 2 z and p{\"o}schl-teller potentials},\ }\href@noop {} {\bibfield  {journal} {\bibinfo  {journal} {Physical Review A}\ }\textbf {\bibinfo {volume} {17}},\ \bibinfo {pages} {1273} (\bibinfo {year} {1978})}\BibitemShut {NoStop}%
\bibitem [{\citenamefont {Petkovi\ifmmode~\acute{c}\else \'{c}\fi{}}\ \emph {et~al.}(2020)\citenamefont {Petkovi\ifmmode~\acute{c}\else \'{c}\fi{}}, \citenamefont {Reichert},\ and\ \citenamefont {Ristivojevic}}]{Petkovic2020}%
  \BibitemOpen
  \bibfield  {author} {\bibinfo {author} {\bibfnamefont {A.}~\bibnamefont {Petkovi\ifmmode~\acute{c}\else \'{c}\fi{}}}, \bibinfo {author} {\bibfnamefont {B.}~\bibnamefont {Reichert}},\ and\ \bibinfo {author} {\bibfnamefont {Z.}~\bibnamefont {Ristivojevic}},\ }\bibfield  {title} {\bibinfo {title} {Density profile of a semi-infinite one-dimensional bose gas and bound states of the impurity},\ }\href {https://doi.org/10.1103/PhysRevResearch.2.043104} {\bibfield  {journal} {\bibinfo  {journal} {Phys. Rev. Res.}\ }\textbf {\bibinfo {volume} {2}},\ \bibinfo {pages} {043104} (\bibinfo {year} {2020})}\BibitemShut {NoStop}%
\bibitem [{\citenamefont {Parisi}\ and\ \citenamefont {Giorgini}(2017)}]{Parisi2017a}%
  \BibitemOpen
  \bibfield  {author} {\bibinfo {author} {\bibfnamefont {L.}~\bibnamefont {Parisi}}\ and\ \bibinfo {author} {\bibfnamefont {S.}~\bibnamefont {Giorgini}},\ }\bibfield  {title} {\bibinfo {title} {Quantum monte carlo study of the bose-polaron problem in a one-dimensional gas with contact interactions},\ }\href {https://doi.org/10.1103/PhysRevA.95.023619} {\bibfield  {journal} {\bibinfo  {journal} {Phys. Rev. A}\ }\textbf {\bibinfo {volume} {95}},\ \bibinfo {pages} {023619} (\bibinfo {year} {2017})}\BibitemShut {NoStop}%
\bibitem [{\citenamefont {Cheon}\ and\ \citenamefont {Shigehara}(1999)}]{cheon1999fermion}%
  \BibitemOpen
  \bibfield  {author} {\bibinfo {author} {\bibfnamefont {T.}~\bibnamefont {Cheon}}\ and\ \bibinfo {author} {\bibfnamefont {T.}~\bibnamefont {Shigehara}},\ }\bibfield  {title} {\bibinfo {title} {Fermion-boson duality of one-dimensional quantum particles with generalized contact interactions},\ }\href@noop {} {\bibfield  {journal} {\bibinfo  {journal} {Physical review letters}\ }\textbf {\bibinfo {volume} {82}},\ \bibinfo {pages} {2536} (\bibinfo {year} {1999})}\BibitemShut {NoStop}%
\bibitem [{\citenamefont {Alexandrov}\ and\ \citenamefont {Krebs}(1992)}]{Alexandrov1992}%
  \BibitemOpen
  \bibfield  {author} {\bibinfo {author} {\bibfnamefont {A.~S.}\ \bibnamefont {Alexandrov}}\ and\ \bibinfo {author} {\bibfnamefont {A.~B.}\ \bibnamefont {Krebs}},\ }\bibfield  {title} {\bibinfo {title} {Polarons in high-temperature superconductors},\ }\href {https://doi.org/10.1070/PU1992v035n05ABEH002235} {\bibfield  {journal} {\bibinfo  {journal} {Phys. Usp.}\ }\textbf {\bibinfo {volume} {35}},\ \bibinfo {pages} {345} (\bibinfo {year} {1992})}\BibitemShut {NoStop}%
\bibitem [{\citenamefont {Mott}(1993)}]{Mott1993}%
  \BibitemOpen
  \bibfield  {author} {\bibinfo {author} {\bibfnamefont {N.}~\bibnamefont {Mott}},\ }\bibfield  {title} {\bibinfo {title} {{Polaron models of high-temperature superconductivity}},\ }\href {https://doi.org/10.1016/0921-4534(93)90187-U} {\bibfield  {journal} {\bibinfo  {journal} {Phys. C Supercond.}\ }\textbf {\bibinfo {volume} {205}},\ \bibinfo {pages} {191} (\bibinfo {year} {1993})}\BibitemShut {NoStop}%
\bibitem [{\citenamefont {Alexandrov}\ and\ \citenamefont {Mott}(1994)}]{Alexandrov1994}%
  \BibitemOpen
  \bibfield  {author} {\bibinfo {author} {\bibfnamefont {A.~S.}\ \bibnamefont {Alexandrov}}\ and\ \bibinfo {author} {\bibfnamefont {N.~F.}\ \bibnamefont {Mott}},\ }\bibfield  {title} {\bibinfo {title} {Bipolarons},\ }\href {https://doi.org/10.1088/0034-4885/57/12/001} {\bibfield  {journal} {\bibinfo  {journal} {Reports on Progress in Physics}\ }\textbf {\bibinfo {volume} {57}},\ \bibinfo {pages} {1197} (\bibinfo {year} {1994})}\BibitemShut {NoStop}%
\bibitem [{\citenamefont {Bredas}\ and\ \citenamefont {Street}(1985)}]{Bredas1985}%
  \BibitemOpen
  \bibfield  {author} {\bibinfo {author} {\bibfnamefont {J.~L.}\ \bibnamefont {Bredas}}\ and\ \bibinfo {author} {\bibfnamefont {G.~B.}\ \bibnamefont {Street}},\ }\bibfield  {title} {\bibinfo {title} {Polarons, bipolarons, and solitons in conducting polymers},\ }\href {https://doi.org/10.1021/ar00118a005} {\bibfield  {journal} {\bibinfo  {journal} {Accounts of Chemical Research}\ }\textbf {\bibinfo {volume} {18}},\ \bibinfo {pages} {309} (\bibinfo {year} {1985})}\BibitemShut {NoStop}%
\bibitem [{\citenamefont {Glenis}\ \emph {et~al.}(1993)\citenamefont {Glenis}, \citenamefont {Benz}, \citenamefont {LeGoff}, \citenamefont {Schindler}, \citenamefont {Kannewurf},\ and\ \citenamefont {Kanatzidis}}]{Glenis1993}%
  \BibitemOpen
  \bibfield  {author} {\bibinfo {author} {\bibfnamefont {S.}~\bibnamefont {Glenis}}, \bibinfo {author} {\bibfnamefont {M.}~\bibnamefont {Benz}}, \bibinfo {author} {\bibfnamefont {E.}~\bibnamefont {LeGoff}}, \bibinfo {author} {\bibfnamefont {J.~L.}\ \bibnamefont {Schindler}}, \bibinfo {author} {\bibfnamefont {C.~R.}\ \bibnamefont {Kannewurf}},\ and\ \bibinfo {author} {\bibfnamefont {M.~G.}\ \bibnamefont {Kanatzidis}},\ }\bibfield  {title} {\bibinfo {title} {Polyfuran: a new synthetic approach and electronic properties},\ }\href {https://doi.org/10.1021/ja00079a035} {\bibfield  {journal} {\bibinfo  {journal} {Journal of the American Chemical Society}\ }\textbf {\bibinfo {volume} {115}},\ \bibinfo {pages} {12519} (\bibinfo {year} {1993})}\BibitemShut {NoStop}%
\bibitem [{\citenamefont {Bussac}\ and\ \citenamefont {Zuppiroli}(1993)}]{Bussac1993}%
  \BibitemOpen
  \bibfield  {author} {\bibinfo {author} {\bibfnamefont {M.~N.}\ \bibnamefont {Bussac}}\ and\ \bibinfo {author} {\bibfnamefont {L.}~\bibnamefont {Zuppiroli}},\ }\bibfield  {title} {\bibinfo {title} {Bipolaron singlet and triplet states in disordered conducting polymers},\ }\href {https://doi.org/10.1103/PhysRevB.47.5493} {\bibfield  {journal} {\bibinfo  {journal} {Phys. Rev. B}\ }\textbf {\bibinfo {volume} {47}},\ \bibinfo {pages} {5493} (\bibinfo {year} {1993})}\BibitemShut {NoStop}%
\bibitem [{\citenamefont {Fernandes}\ \emph {et~al.}(2005)\citenamefont {Fernandes}, \citenamefont {Garcia}, \citenamefont {Schultz},\ and\ \citenamefont {Nart}}]{Fernandes2005}%
  \BibitemOpen
  \bibfield  {author} {\bibinfo {author} {\bibfnamefont {M.}~\bibnamefont {Fernandes}}, \bibinfo {author} {\bibfnamefont {J.}~\bibnamefont {Garcia}}, \bibinfo {author} {\bibfnamefont {M.}~\bibnamefont {Schultz}},\ and\ \bibinfo {author} {\bibfnamefont {F.}~\bibnamefont {Nart}},\ }\bibfield  {title} {\bibinfo {title} {Polaron and bipolaron transitions in doped poly(p-phenylene vinylene) films},\ }\href {https://doi.org/https://doi.org/10.1016/j.tsf.2004.08.066} {\bibfield  {journal} {\bibinfo  {journal} {Thin Solid Films}\ }\textbf {\bibinfo {volume} {474}},\ \bibinfo {pages} {279 } (\bibinfo {year} {2005})}\BibitemShut {NoStop}%
\bibitem [{\citenamefont {Zozoulenko}\ \emph {et~al.}(2019)\citenamefont {Zozoulenko}, \citenamefont {Singh}, \citenamefont {Singh}, \citenamefont {Gueskine}, \citenamefont {Crispin},\ and\ \citenamefont {Berggren}}]{Zozoulenko2019}%
  \BibitemOpen
  \bibfield  {author} {\bibinfo {author} {\bibfnamefont {I.}~\bibnamefont {Zozoulenko}}, \bibinfo {author} {\bibfnamefont {A.}~\bibnamefont {Singh}}, \bibinfo {author} {\bibfnamefont {S.~K.}\ \bibnamefont {Singh}}, \bibinfo {author} {\bibfnamefont {V.}~\bibnamefont {Gueskine}}, \bibinfo {author} {\bibfnamefont {X.}~\bibnamefont {Crispin}},\ and\ \bibinfo {author} {\bibfnamefont {M.}~\bibnamefont {Berggren}},\ }\bibfield  {title} {\bibinfo {title} {Polarons, bipolarons, and absorption spectroscopy of pedot},\ }\href {https://doi.org/10.1021/acsapm.8b00061} {\bibfield  {journal} {\bibinfo  {journal} {ACS Applied Polymer Materials}\ }\textbf {\bibinfo {volume} {1}},\ \bibinfo {pages} {83} (\bibinfo {year} {2019})}\BibitemShut {NoStop}%
\bibitem [{\citenamefont {Bobbert}\ \emph {et~al.}(2007)\citenamefont {Bobbert}, \citenamefont {Nguyen}, \citenamefont {van Oost}, \citenamefont {Koopmans},\ and\ \citenamefont {Wohlgenannt}}]{Bobbert2007}%
  \BibitemOpen
  \bibfield  {author} {\bibinfo {author} {\bibfnamefont {P.~A.}\ \bibnamefont {Bobbert}}, \bibinfo {author} {\bibfnamefont {T.~D.}\ \bibnamefont {Nguyen}}, \bibinfo {author} {\bibfnamefont {F.~W.~A.}\ \bibnamefont {van Oost}}, \bibinfo {author} {\bibfnamefont {B.}~\bibnamefont {Koopmans}},\ and\ \bibinfo {author} {\bibfnamefont {M.}~\bibnamefont {Wohlgenannt}},\ }\bibfield  {title} {\bibinfo {title} {Bipolaron mechanism for organic magnetoresistance},\ }\href {https://doi.org/10.1103/PhysRevLett.99.216801} {\bibfield  {journal} {\bibinfo  {journal} {Phys. Rev. Lett.}\ }\textbf {\bibinfo {volume} {99}},\ \bibinfo {pages} {216801} (\bibinfo {year} {2007})}\BibitemShut {NoStop}%
\bibitem [{\citenamefont {Schecter}\ \emph {et~al.}(2016)\citenamefont {Schecter}, \citenamefont {Gangardt},\ and\ \citenamefont {Kamenev}}]{Schecter2016}%
  \BibitemOpen
  \bibfield  {author} {\bibinfo {author} {\bibfnamefont {M.}~\bibnamefont {Schecter}}, \bibinfo {author} {\bibfnamefont {D.~M.}\ \bibnamefont {Gangardt}},\ and\ \bibinfo {author} {\bibfnamefont {A.}~\bibnamefont {Kamenev}},\ }\bibfield  {title} {\bibinfo {title} {{Quantum impurities: from mobile Josephson junctions to depletons}},\ }\href {https://doi.org/10.1088/1367-2630/18/6/065002} {\bibfield  {journal} {\bibinfo  {journal} {New J. Phys.}\ }\textbf {\bibinfo {volume} {18}},\ \bibinfo {pages} {065002} (\bibinfo {year} {2016})}\BibitemShut {NoStop}%
\bibitem [{\citenamefont {Reichert}\ \emph {et~al.}(2019)\citenamefont {Reichert}, \citenamefont {Petkovi{\'{c}}},\ and\ \citenamefont {Ristivojevic}}]{Reichert2019}%
  \BibitemOpen
  \bibfield  {author} {\bibinfo {author} {\bibfnamefont {B.}~\bibnamefont {Reichert}}, \bibinfo {author} {\bibfnamefont {A.}~\bibnamefont {Petkovi{\'{c}}}},\ and\ \bibinfo {author} {\bibfnamefont {Z.}~\bibnamefont {Ristivojevic}},\ }\bibfield  {title} {\bibinfo {title} {{Field-theoretical approach to the Casimir-like interaction in a one-dimensional Bose gas}},\ }\href@noop {} {\bibfield  {journal} {\bibinfo  {journal} {Phys. Rev. B}\ }\textbf {\bibinfo {volume} {99}},\ \bibinfo {pages} {205414} (\bibinfo {year} {2019})}\BibitemShut {NoStop}%
\bibitem [{\citenamefont {Petkovi{\'c}}\ and\ \citenamefont {Ristivojevic}(2022)}]{petkovic2022mediated}%
  \BibitemOpen
  \bibfield  {author} {\bibinfo {author} {\bibfnamefont {A.}~\bibnamefont {Petkovi{\'c}}}\ and\ \bibinfo {author} {\bibfnamefont {Z.}~\bibnamefont {Ristivojevic}},\ }\bibfield  {title} {\bibinfo {title} {Mediated interaction between polarons in a one-dimensional bose gas},\ }\href@noop {} {\bibfield  {journal} {\bibinfo  {journal} {Physical Review A}\ }\textbf {\bibinfo {volume} {105}},\ \bibinfo {pages} {L021303} (\bibinfo {year} {2022})}\BibitemShut {NoStop}%
\bibitem [{\citenamefont {Schecter}\ and\ \citenamefont {Kamenev}(2014)}]{schecter2014phonon}%
  \BibitemOpen
  \bibfield  {author} {\bibinfo {author} {\bibfnamefont {M.}~\bibnamefont {Schecter}}\ and\ \bibinfo {author} {\bibfnamefont {A.}~\bibnamefont {Kamenev}},\ }\bibfield  {title} {\bibinfo {title} {Phonon-mediated casimir interaction between mobile impurities in one-dimensional quantum liquids},\ }\href@noop {} {\bibfield  {journal} {\bibinfo  {journal} {Physical review letters}\ }\textbf {\bibinfo {volume} {112}},\ \bibinfo {pages} {155301} (\bibinfo {year} {2014})}\BibitemShut {NoStop}%
\bibitem [{\citenamefont {Huber}\ \emph {et~al.}(2019)\citenamefont {Huber}, \citenamefont {Hammer},\ and\ \citenamefont {Volosniev}}]{huber2019medium}%
  \BibitemOpen
  \bibfield  {author} {\bibinfo {author} {\bibfnamefont {D.}~\bibnamefont {Huber}}, \bibinfo {author} {\bibfnamefont {H.-W.}\ \bibnamefont {Hammer}},\ and\ \bibinfo {author} {\bibfnamefont {A.}~\bibnamefont {Volosniev}},\ }\bibfield  {title} {\bibinfo {title} {In-medium bound states of two bosonic impurities in a one-dimensional fermi gas},\ }\href@noop {} {\bibfield  {journal} {\bibinfo  {journal} {Physical Review Research}\ }\textbf {\bibinfo {volume} {1}},\ \bibinfo {pages} {033177} (\bibinfo {year} {2019})}\BibitemShut {NoStop}%
\bibitem [{\citenamefont {Volosniev}\ \emph {et~al.}(2015)\citenamefont {Volosniev}, \citenamefont {Hammer},\ and\ \citenamefont {Zinner}}]{Volosniev2015}%
  \BibitemOpen
  \bibfield  {author} {\bibinfo {author} {\bibfnamefont {A.~G.}\ \bibnamefont {Volosniev}}, \bibinfo {author} {\bibfnamefont {H.~W.}\ \bibnamefont {Hammer}},\ and\ \bibinfo {author} {\bibfnamefont {N.~T.}\ \bibnamefont {Zinner}},\ }\bibfield  {title} {\bibinfo {title} {{Real-time dynamics of an impurity in an ideal Bose gas in a trap}},\ }\href {https://doi.org/10.1103/PhysRevA.92.023623} {\bibfield  {journal} {\bibinfo  {journal} {Phys. Rev. A - At. Mol. Opt. Phys.}\ }\textbf {\bibinfo {volume} {92}},\ \bibinfo {pages} {023623} (\bibinfo {year} {2015})}\BibitemShut {NoStop}%
\bibitem [{\citenamefont {Dehkharghani}\ \emph {et~al.}(2015)\citenamefont {Dehkharghani}, \citenamefont {Volosniev},\ and\ \citenamefont {Zinner}}]{dehkharghani2015quantum}%
  \BibitemOpen
  \bibfield  {author} {\bibinfo {author} {\bibfnamefont {A.~S.}\ \bibnamefont {Dehkharghani}}, \bibinfo {author} {\bibfnamefont {A.}~\bibnamefont {Volosniev}},\ and\ \bibinfo {author} {\bibfnamefont {N.}~\bibnamefont {Zinner}},\ }\bibfield  {title} {\bibinfo {title} {Quantum impurity in a one-dimensional trapped bose gas},\ }\href@noop {} {\bibfield  {journal} {\bibinfo  {journal} {Physical Review A}\ }\textbf {\bibinfo {volume} {92}},\ \bibinfo {pages} {031601} (\bibinfo {year} {2015})}\BibitemShut {NoStop}%
\bibitem [{\citenamefont {G{\'o}mez-Lozada}\ \emph {et~al.}(2024)\citenamefont {G{\'o}mez-Lozada}, \citenamefont {Hiyane}, \citenamefont {Busch},\ and\ \citenamefont {Fogarty}}]{gomez2024bose}%
  \BibitemOpen
  \bibfield  {author} {\bibinfo {author} {\bibfnamefont {F.}~\bibnamefont {G{\'o}mez-Lozada}}, \bibinfo {author} {\bibfnamefont {H.}~\bibnamefont {Hiyane}}, \bibinfo {author} {\bibfnamefont {T.}~\bibnamefont {Busch}},\ and\ \bibinfo {author} {\bibfnamefont {T.}~\bibnamefont {Fogarty}},\ }\bibfield  {title} {\bibinfo {title} {Bose-fermi $ n $-polaron state emergence from correlation-mediated blocking of phase separation},\ }\href@noop {} {\bibfield  {journal} {\bibinfo  {journal} {arXiv preprint arXiv:2409.13785}\ } (\bibinfo {year} {2024})}\BibitemShut {NoStop}%
\bibitem [{\citenamefont {Lawden}(1989)}]{Lawden1989}%
  \BibitemOpen
  \bibfield  {author} {\bibinfo {author} {\bibfnamefont {D.~F.}\ \bibnamefont {Lawden}},\ }\href@noop {} {\emph {\bibinfo {title} {Elliptic Functions and Applications}}}\ (\bibinfo  {publisher} {Applied Mathematical Sciences},\ \bibinfo {year} {1989})\BibitemShut {NoStop}%
\end{thebibliography}%

\end{document}